\documentclass[doublespacing]{bmcart}

\usepackage{amsthm,amsmath}
\usepackage{url}
\RequirePackage{natbib}
\RequirePackage{hyperref}
\usepackage[utf8]{inputenc} %unicode support
\usepackage[T1]{fontenc}
\usepackage{bbm}
\usepackage{amssymb}
\usepackage{xifthen}
\usepackage{tikz}
\usepackage{xargs}
\usepackage[colorinlistoftodos,prependcaption,textsize=tiny]{todonotes}
\newcommandx{\unsure}[2][1=]{\todo[linecolor=red,backgroundcolor=red!25,bordercolor=red,#1]{#2}}
\newcommandx{\change}[2][1=]{\todo[linecolor=blue,backgroundcolor=blue!25,bordercolor=blue,#1]{#2}}
\newcommandx{\info}[2][1=]{\todo[linecolor=green,backgroundcolor=green!25,bordercolor=green,#1]{#2}}
\newcommandx{\thiswillnotshow}[2][1=]{\todo[disable,#1]{#2}}

\graphicspath{{./../figures/}}

\startlocaldefs

\definecolor{petrol}{RGB}{57,119,132}

\newcommand{\larg}{\mathopen{}\mathclose\bgroup\left} % \left command for function arguments (correct spacing, '(' 
\newcommand{\rarg}{\aftergroup\egroup\right} % same for the \right
\newcommand{\fargs}[1]{\larg(#1\rarg)}
\newcommand{\one}{\mathbbm{1}}
\newcommand{\set}[1]{\left\{#1\right\}} % a set
\DeclareMathOperator{\diag}{\mathrm{diag}} % diagonal
\newcommand{\diagm}[1]{\diag\larg\{#1\rarg\}} % diagonal matrix
\newcommand{\ci}{\mathrm{i}} % complex i 
\DeclareMathOperator{\myRe}{\mathrm{Re}} % Real part
\DeclareMathOperator{\myIm}{\mathrm{Im}} % Complex part
\DeclareMathOperator{\ind}{\mathrm{ind}}
\DeclareMathOperator{\res}{\mathrm{res}} % residue operator
\newcommand{\residue}[2]{\res_{#1}{#2}} % residue

\newcommand{\diffd}{\mathrm{d}} 

\newcommand{\intd}{\,\mathrm{d}} 
\newcommand{\PId}{\,\mathcal{D}} 

\newcommand{\prob}[1]{\mathcal{P}\larg[#1\rarg]} % probability
\DeclareMathOperator{\covarop}{\mathrm{Cov}} % variance operator
\newcommand{\covariance}[2]{\covarop\larg[#1,#2\rarg]} % covariance
\newcommand{\ensexpect}[1]{\left\langle #1 \right\rangle} % ensemble expectation
\newcommand{\Poiss}{\operatorname{\mathrm{Poiss}}}

\newcommand{\thetafct}[1]{\Theta\larg(#1\rarg)} % Theta function
\newcommand{\deltafct}[1]{\delta\larg(#1\rarg)} % delta distribution
\DeclareMathOperator{\smallo}{\mathrm{o}} % small o notation
\newcommand{\conv}[2]{\larg(#1\star#2\rarg)}

\newcommand{\mdef}{=}

\newcommand{\FT}[1]{\widehat{#1}}
\newcommand{\dt}{\Delta t}
\newcommand{\dprime}{{\prime\prime}}

\newcommand{\ER}{Erd\H{o}s--R\'{e}nyi}

\newcommand{\Lloopidx}{{k_{M}\dots k_{0},l_{N}\dots l_{0}}}
\newcommand{\Iloopidx}{{i_0\dots i_{M-1},j_0\dots j_{N-1}}}
\newcommand{\De}{D^\mathrm{e}}
\newcommand{\Da}{D^\alpha}

\newcommand{\convfirst}[2]{\larg(#1\star_1#2\rarg)}

\newcommand{\treeprop}[1]{\ifthenelse{\isempty{#1}}{\Delta_T}{\Delta_{T,#1}}} % tree-level propagator (with optional index args)
\newcommand{\FTtreeprop}[1]{\ifthenelse{\isempty{#1}}{\FT{\Delta}_T}{\FT{\Delta}_{T,#1}}} % Fourier-transformed tree-level propagator (with optional args)
\newcommand{\invtreeprop}[1]{\ifthenelse{\isempty{#1}}{\Delta_T^{-1}}{\ensuremath{\Delta_{T,#1}^{-1}}}} % inverse tree-level propagator (with optional args)

\newcommand{\actarg}{\larg[\tilde z, z\rarg]} % argument of the (non-shifted) action
\newcommand{\effarg}{\larg[\tilde z,\delta z\rarg]} % argument of the effective action

\newcommand{\treeexpect}[1]{\ensexpect{#1}_T} % tree-level expectation
\newcommand{\freeexpect}[1]{\ensexpect{#1}_F} % expectation according to free action

\usetikzlibrary{matrix,arrows,calc,decorations,decorations.markings,decorations.pathmorphing,positioning}

\tikzstyle{outervertexstyle}=[draw,circle,black,bottom color=white,
                  top color= white, text=black,inner sep=0pt,minimum size=5pt]
\tikzstyle{innervertexstyle}=[draw,circle,black,bottom color=black,
                  top color= black, text=white,inner sep=0pt,minimum size=5pt]
\tikzstyle{arrowedge} = [draw,thick,postaction={decorate,decoration={
    markings,
    mark=at position 0.5 with {\arrow[scale=1]{>}}}}]
\tikzstyle{dottededge} = [draw,loosely dotted,thick]
\tikzstyle{filteredge} = [draw,thick,decorate,decoration={snake}]

\newcommand{\Hz}{\ensuremath{\,\mathrm{Hz}}}
\newcommand{\millisec}{\ensuremath{\,\textrm{ms}}}

\endlocaldefs

\begin{document}

%%% Start of article front matter
\begin{frontmatter}

\begin{fmbox}
\dochead{Research}


\title{Spike Train Cumulants for Linear-Nonlinear Poisson Cascade Models}

%%%%%%%%%%%%%%%%%%%%%%%%%%%%%%%%%%%%%%%%%%%%%%
%%                                          %%
%% Enter the authors here                   %%
%%                                          %%
%% Specify information, if available,       %%
%% in the form:                             %%
%%   <key>={<id1>,<id2>}                    %%
%%   <key>=                                 %%
%% Comment or delete the keys which are     %%
%% not used. Repeat \author command as much %%
%% as required.                             %%
%%                                          %%
%%%%%%%%%%%%%%%%%%%%%%%%%%%%%%%%%%%%%%%%%%%%%%

\author[
   addressref={aff1},                   % id's of addresses, e.g. {aff1,aff2}
%    corref={aff1},                       % id of corresponding address, if any
%    noteref={n1},                        % id's of article notes, if any
%    email={michael.kordovan@bcf.uni-freiburg.de}   % email address
]{\inits{MK}\fnm{Michael} \snm{Kordovan}}
\author[
   addressref={aff1},
   corref={aff1},                       % id of corresponding address, if any
   email={stefan.rotter@bio.uni-freiburg.de}
]{\inits{SR}\fnm{Stefan} \snm{Rotter}}

%%%%%%%%%%%%%%%%%%%%%%%%%%%%%%%%%%%%%%%%%%%%%%
%%                                          %%
%% Enter the authors' addresses here        %%
%%                                          %%
%% Repeat \address commands as much as      %%
%% required.                                %%
%%                                          %%
%%%%%%%%%%%%%%%%%%%%%%%%%%%%%%%%%%%%%%%%%%%%%%

\address[id=aff1]{%                           % unique id
  \orgname{Bernstein Center Freiburg \& Faculty of Biology, University of Freiburg}, % university, etc
  \street{Hansastraße 9a},                     %
  \postcode{79104}                            % post or zip code
  \city{Freiburg},                            % city
  \cny{Germany}                               % country
}

%%%%%%%%%%%%%%%%%%%%%%%%%%%%%%%%%%%%%%%%%%%%%%
%%                                          %%
%% Enter short notes here                   %%
%%                                          %%
%% Short notes will be after addresses      %%
%% on first page.                           %%
%%                                          %%
%%%%%%%%%%%%%%%%%%%%%%%%%%%%%%%%%%%%%%%%%%%%%%

\begin{artnotes}
%\note{Sample of title note}     % note to the article
% \note[id=n1]{Equal contributor} % note, connected to author
\end{artnotes}

\end{fmbox}% comment this for two column layout

%%%%%%%%%%%%%%%%%%%%%%%%%%%%%%%%%%%%%%%%%%%%%%
%%                                          %%
%% The Abstract begins here                 %%
%%                                          %%
%% Please refer to the Instructions for     %%
%% authors on http://www.biomedcentral.com  %%
%% and include the section headings         %%
%% accordingly for your article type.       %%
%%                                          %%
%%%%%%%%%%%%%%%%%%%%%%%%%%%%%%%%%%%%%%%%%%%%%%

\begin{abstractbox}

\begin{abstract} % abstract
% Intro
Spiking activity in cortical networks is nonlinear in nature. 
The linear-nonlinear cascade model, some versions of which are also known as point-process generalized linear model,
can efficiently capture the nonlinear dynamics exhibited by such networks.
Of particular interest in such models are theoretical predictions of spike train statistics. 
However, due to the moment-closure problem, approximations are inevitable.
We suggest here a series expansion that explains how higher-order moments couple to lower-order ones.
% Objective
Our approach makes predictions in terms of certain integrals, the so-called loop integrals.
In previous studies these integrals have been evaluated numerically, but numerical instabilities are sometimes encountered rendering the results unreliable. 
Analytic solutions are presented here to overcome this problem, and to arrive at more robust evaluations.
% Methods
We were able to deduce these analytic solutions by switching to Fourier space and making use of complex analysis, specifically Cauchy's residue theorem.
% Results
We formalized the loop integrals and explicitly solved them for specific response functions. 
To quantify the importance of these corrections for spike train cumulants, we numerically simulated spiking networks and compared their sample statistics to our theoretical predictions.
Our results demonstrate that the magnitude of the nonlinear corrections depends on the working point of the nonlinear network dynamics, and that it is related to the eigenvalues of the mean-field stability matrix.
For our example, the corrections for the firing rates are in the range between $4\,\%$ and $21\,\%$ on average.
% Conclusions
Precise and robust predictions of spike train statistics accounting for nonlinear effects
are, for example, highly relevant for theories involving spike-timing dependent plasticity (STDP).
\end{abstract}

%%%%%%%%%%%%%%%%%%%%%%%%%%%%%%%%%%%%%%%%%%%%%%
%%                                          %%
%% The keywords begin here                  %%
%%                                          %%
%% Put each keyword in separate \kwd{}.     %%
%%                                          %%
%%%%%%%%%%%%%%%%%%%%%%%%%%%%%%%%%%%%%%%%%%%%%%

\begin{keyword}
\kwd{Spike train cumulants}
\kwd{Linear-nonlinear Poisson model}
\kwd{Path integral formalism}
\kwd{Structure-dynamics relations}
\end{keyword}

% MSC classifications codes, if any
%\begin{keyword}[class=AMS]
%\kwd[Primary ]{}
%\kwd{}
%\kwd[; secondary ]{}
%\end{keyword}

\end{abstractbox}
%
%\end{fmbox}% uncomment this for twcolumn layout

\end{frontmatter}

%%%%%%%%%%%%%%%%%%%%%%%%%%%%%%%%%%%%%%%%%%%%%%
%%                                          %%
%% The Main Body begins here                %%
%%                                          %%
%% Please refer to the instructions for     %%
%% authors on:                              %%
%% http://www.biomedcentral.com/info/authors%%
%% and include the section headings         %%
%% accordingly for your article type.       %%
%%                                          %%
%% See the Results and Discussion section   %%
%% for details on how to create sub-sections%%
%%                                          %%
%% use \cite{...} to cite references        %%
%%  \cite{koon} and                         %%
%%  \cite{oreg,khar,zvai,xjon,schn,pond}    %%
%%  \nocite{smith,marg,hunn,advi,koha,mouse}%%
%%                                          %%
%%%%%%%%%%%%%%%%%%%%%%%%%%%%%%%%%%%%%%%%%%%%%%

%%%%%%%%%%%%%%%%%%%%%%%%% start of article main body
% <put your article body there>

%%%%%%%%%%%%%%%%%%%%%%
\section{Introduction}
%%%%%%%%%%%%%%%%%%%%%%
%%%%%%%%%%%%%%%%%%%%%%

Novel experimental techniques for neuronal recordings generate 
huge volumes of data. These data contain information about neuronal spike trains, 
the firing rate of individual neurons, and correlations between neurons, but also information about their connectivity \cite{barthoCharacterizationNeocorticalPrincipal2004, peyracheSpatiotemporalDynamicsNeocortical2012,lee_anatomy_2016,kleinfeld_large-scale_2011,takemura_visual_2013,helmstaedter_connectomic_2013,briggman_wiring_2011,bock_network_2011,kasthuri_saturated_2015,mishchenko_ultrastructural_2010}.

Modeling neuronal activity can be performed at different scales \cite{Herz_2006}. 
The largest amount of neurophysiological detail is conveyed by the simulation of neuron models with dendrites and axons extending in three-dimensional space \cite{Koch_book_2001}.
Apart from the problem that most of the details concerning neuron morphology and ion channel equipment are not known, this approach is computationally very demanding, and may become entirely unfeasible for larger networks of neurons. 
Consequently, most studies of large-scale spiking network dynamics use simpler point neuron models. 
They come as deterministic or stochastic units. A well-known example of the former type is the \emph{leaky integrate-and-fire neuron} \cite{stein_1965,brunel_2000}, a prominent representative of the
latter are self-exciting and mutually-exciting Poisson processes, called \emph{Hawkes processes} \cite{hawkesA_1971,hawkesB_1971}.
Studying biophysically inspired models and abstract point processes side-by-side can be of great help to understand the influence of network structure on spike train statistics in recurrent networks \cite{Pernice2011, Pernice2012, Jovanovic_2015, Jovanovic_PLoSCB2016}. % remark: Trousdale2012 and Hu2013 do not consider Hawkes processes, not mention here

As the classical Hawkes model is fully linear, the nonlinear dynamics of biological neuronal networks must be linearized before comparing them. This poses limits to the range of applications of this model, and to the precision of the results achieved with it.
A natural generalization emerges by including an arbitrary transfer function for the firing rates to account for intrinsic neuronal nonlinearities.
Such a model, called \emph{linear-nonlinear Poisson (LNP) cascade model}, naturally emerges for several spiking neuron models \cite{ostojic_spikingtoLNP_2011}.
As a phenomenological model, it has also been successfully employed for the analysis of multiple-neuron spike trains \cite{Truccolo2016,Pillow2008,gerhard_PPGLM_2017,lambert_reconstructingconnsHawkes_2017}.

Computing spike train cumulants of any order in the LNP model is a daunting task.
However, the precise knowledge of certain low-order cumulants is of great importance for the study of spike-timing dependent plasticity (STDP) \cite{Feldman_STDPreview_2012}.
For simple pairwise STDP, first- and second-order cumulants are sufficient, but the more complex model of triplet STDP \cite{pfister_triplets_2006} requires accurate knowledge of third-order spike train cumulants. 
Furthermore, there is evidence that higher-order cumulants constrain neuronal activity patterns, and it has been shown that including them into statistical models improves the fit of experimental data \cite{shimazaki_HOI_2015}. 
For predictions in nonlinear models, the coupling of higher-order moments to lower-order ones complicates the calculations. This issue is known as the \emph{moment-closure problem}.
One systematic way to deal with this problem and to manage the complexity of the hierarchy of contributions has been proposed by \cite{Buice2010} in terms of a path integral formulation.
The authors in \cite{GKO_PILNP_2017} used this method to calculate perturbative corrections to the
predictions made by the linear model. 
The corrections arise as higher-order moments couple to lower-order ones, due to the nonlinearity. 

In principle, the problem might be considered to be solved, and in theory it is. But when it comes to actual numerical predictions of cumulants, several technical difficulties arise.
In the diagrammatic expansion, corrections stemming from the nonlinear transfer functions involve diagrams with loops. 
When calculating the contributions of these loop diagrams, integrals over the loop momentum have to be solved. 
Only analytic solutions guarantee a correct solution of these integrals, independent of choices made for the parameters. 
Here we demonstrate how loop integrals can be analytically solved using methods from complex analysis.
For certain parameter regimes numerical estimates can be misleading (or wrong) and analytic solutions are preferred.

%%%%%%%%%%%%%%%%%%%%%%%%%%%%%%%%%%%
\section{Spiking Model and Methods}
%%%%%%%%%%%%%%%%%%%%%%%%%%%%%%%%%%%
%%%%%%%%%%%%%%%%%%%%%%%%%%%%%%%%%%%

In a network of spiking units, the individual node activities generally depend on different variables. 
These variables reflect the applied stimulus, previous history of activity, and neuronal coupling \cite{Kass_book_2014}. 
We first review a widely used stochastic model, and then outline a less well-known representation of it \cite{GKO_PILNP_2017}. 

\subsection{Linear-nonlinear Poisson cascade model}

Dynamics in neuronal networks can be mathematically described as stochastic point processes \cite{GruenRotter_book_2010}. 
The spikes associated to a neuron in a network correspond to discrete events in time.
Individual spikes are random but not necessarily stochastically independent \cite{GruenRotter_book_2010,Pillow2008,lambert_reconstructingconnsHawkes_2017}. 

A point process can be defined by use of discrete event times, inter-event intervals, or a cumulative counting process \cite{CoxIsham_PPbook_1980,Daley_PPv1_2003}. 
A \emph{point process} is a random sequence $\mathcal{T} = \left[t^\gamma\right]_{\gamma\geq 1}$ with $t^0=0$, $t^\gamma\in\left[0,\infty\right)$ and $t^\gamma<t^{\gamma+1}$ \cite{Daley_PPv1_2003}.
A useful representation of a point process is given by the collection of all events,
\begin{equation}\label{eq:RCMrep}
 z\fargs{t}=\sum_{t^\gamma} \deltafct{t-t^\gamma},
\end{equation}
where $\deltafct{t}$ is the Dirac measure. 
The associated counting process $N\fargs{t}$ is given by 
\begin{equation}
 N\fargs{t}=\int_0^t z\fargs{t}\intd t.
\end{equation}
If $\prob{N(t,t+x]=k}$ only depends on the duration $x$, but not on the location $t$, the point process is called \emph{crudely stationary} \cite{Daley_PPv1_2003}. 
For crudely stationary point processes, Khinchin's Existence Theorem \cite[Proposition~3.3.I]{Daley_PPv1_2003} guarantees the existence of the intensity
\begin{equation}
 \lambda = \lim_{\diffd t\searrow 0} \frac{\prob{N(0,t]>0}}{\diffd t},
\end{equation}
although it might be infinite. In case of finite intensities, it is meaningful to write
\begin{equation}
 \prob{N(t,t+\diffd t] > 0} = \lambda\diffd t +\smallo\fargs{\diffd t}.
\end{equation}
Further, a point process is said to be \emph{simple} \cite{Daley_PPv1_2003}, if
\begin{equation}
 \prob{N\fargs{\set{t}} = 0 \text{ or } 1~ \forall t} = 1.
\end{equation}
A crudely stationary point process is \emph{orderly} \cite{Daley_PPv1_2003}, when
\begin{equation}
 \prob{N(0,\diffd t]\geq 2} = \smallo\fargs{\diffd t},\qquad(\diffd t\searrow 0).
\end{equation}
For crudely stationary point processes of finite intensity, these to properties are equivalent \cite{Daley_PPv1_2003}.
The most prominent example is a stationary Poisson process, completely defined by \cite{Daley_PPv1_2003}
\begin{equation}
 \prob{N(a_i,b_i]=n_i,i=1,\dots,k}=\prod_{i=1}^k \frac{\left(\lambda\left(b_i-a_i\right)\right)^{n_i}}{n_i!} e^{-\lambda\left(b_i-a_i\right)} ,
\end{equation}
with $a_i<b_i\leq a_{i+1}$.

In a multivariate setting, each neuron $i$ has an associated point process $\mathcal{T}_i$ describing its spike times. The corresponding point process representation $z_i\fargs{t}$ from Eq.~\eqref{eq:RCMrep} is called \emph{spike train}.
The \emph{spike count} of neuron $i$ is the counting process $N_i\fargs{t}$ associated with $\mathcal{T}_i$. It is defined by 
\begin{equation}
 N_i\fargs{t} = \sum_{\gamma=1}^{\infty}\thetafct{t - t^\gamma_i},
\end{equation}
or equivalently 
\begin{equation}\label{eq:CPrep}
 N_i\fargs{t} = \int_{0}^t z_i\fargs{t}\intd t.
\end{equation}

Assuming orderliness, a multivariate counting Process ${\left(N_i\fargs{t},\dots,N_n\fargs{t}\right)_{t\geq 0}}$ is entirely characterized by its conditional intensity process $\left(\lambda_i\fargs{t},\dots,\lambda_n\fargs{t}\right)_{t\geq 0}$. For finite intensity processes we may write 
\begin{equation}
 \prob{\diffd N_i\fargs{t}=1|\mathcal{H}_t}=\lambda_i\fargs{t}\intd t,
\end{equation}
where $\mathcal{H}_t$ is the history of the point process up to time $t$ \cite{CoxIsham_PPbook_1980}.

An interesting candidate is given by linear-nonlinear cascade models \cite{Paninski_LNP_2004,Simoncelli2004, ostojic_spikingtoLNP_2011},
where the intensities read
\begin{equation}
 \lambda_i\fargs{t} = \phi_i\fargs{\sum_j\int_{-\infty}^{t-}h_{ij}\fargs{t-s}\intd N_j\fargs{s} + b_i\fargs{t}}.
\end{equation}
Here, $h_{ij}$ is the (causal) response function describing the influence that neuron $j$ exerts on neuron $i$. The nonnegative gain function $\phi_i$ accounts for the nonlinear transfer from the intrinsic state (``membrane potential'') to the intensity (``firing rate''), and $b_i\fargs{t}$ specifies the baseline intensity.
The linear filtering can be thought of as the spatiotemporal receptive field of the neuron under investigation \cite{Paninski_LNP_2004}. By use of Eq.~\eqref{eq:CPrep}, we have
\begin{equation}\label{eq:intensity_LNP}
 \lambda_i\fargs{t} = \phi_i\fargs{\sum_j\conv{h_{ij}}{z_j}\fargs{t} +b_i\fargs{t}},
\end{equation}
where $\conv{f}{g}\fargs{t}$ denotes the usual convolution operation. Integration boundaries and causality are reflected by an appropriate definition of $h_{ij}$.

The connectivity in a neuronal network is specified by the adjacency matrix $A$ 
with entries $a_{ij}\in \set{0,1}$, or $a_{ij}\in\mathbbm{N}_0$ if multiple connections are allowed between two neurons. 
As the effects on post-synaptic neurons can vastly differ in magnitude, and synapses can be either excitatory or inhibitory, synapses carry 
a signed weight $w_{ij}$. The connectivity matrix $W$ is given by a Hadamard product of adjacency matrix and weight matrix.
Finally, any time-dependencies of the neuronal influence are reflected by the synaptic kernel functions. They might vary for different neuron types and thus, the matrix of causal interaction filters $H=\left(h_{ij}\right)_{i,j\in\set{1,\dots,n}}$ is the Hadamard product of the matrix of response functions and the connectivity matrix. 

We term this model \emph{linear-nonlinear Poisson (LNP) cascade model} similar to \cite{Simoncelli2004}. 
Poisson means that spikes are drawn from a Poisson distribution with instantaneous rate $\lambda$,
but due to the self-interaction and history dependence the output spike trains do clearly not fulfill the independence criteria of an inhomogeneous Poisson process any more.
These processes are further known as (nonlinear) Hawkes processes if all interactions are non-negative \cite{Bremaud1996,Daley_PPv1_2003,Delattre2016}.

% \subsection{Related Nonlinear Models}
A different formulation which is better suited for estimation of these processes is obtained by means of the point process generalized linear model framework \cite{Truccolo2005,Pillow2008}.
More specifically, the joint probability density of the process is approximated by a discrete one.
This defines a likelihood function belonging to the exponential family with canonical parameter $\log\lambda_i\fargs{t}$ \cite{Truccolo2005,Truccolo2016}. 
In the generalized linear model setting this canonical parameter is expressed as linear combination of covariates \cite{Nelder1972_GLM}.
Thus, the LNP cascade model with exponential gain function can directly be obtained \cite{Truccolo2016,gerhard_PPGLM_2017}.
The exponential gain function further implies multiplicative effects from previous spikes on the instantaneous firing rates \cite{Cardanobile2010}.
Point process generalized linear models are very promising and successful models for spike responses of single neurons or networks 
\cite{Paninski_LNP_2004,Truccolo2005,Pillow2008,Kass_book_2014,weber_GLM_2017}.

A concluding remark concerns the nonlinear gain function. Nonlinearities can enter at two different levels, pre-synaptic or 
post-synaptic, corresponding to the connectivity matrix appearing outside or inside the nonlinearity.
For the former, an Amari-type model \cite{Amari1975,Amari1977}, the rate is given by
\begin{equation}
 \lambda_i\larg(t\rarg) = \kappa \sum_j W_{ij}\phi\larg(z_j\larg(t\rarg)\rarg)+b_i\larg(t\rarg).
\end{equation}
Both models imply different corrections to mean-field equations as pointed out by \cite{Bressloff2015}.
A further possibility to incorporate nonlinearities into the model is by considering Volterra series expansions. Recent progress in this direction can be found in \cite{Voronenko_thesis_2018}.

\subsection{Path integral representation of the linear-nonlinear Poisson cascade model}

Since the moment generating functional for the LNP cascade model cannot be calculated analytically, the following alternative strategy is applied. First, an auxiliary variable, the so-called \emph{response variable}, is introduced. Its dynamics describes the stochastic evolution of the system under consideration \cite{GKO_PILNP_2017}. Then, the probability density functional (pdf) of the process is written in exponential form, where the negative exponent is called the \emph{action}. The action splits into the \emph{free action}, which is bilinear in the configuration and the response variable, and the \emph{interaction} component, which comprises all remaining terms. While expectation values with respect to the pdf of the free action can be calculated, this is generally not possible for the full action. However, the interaction component can be expanded into a series. Finally, the moments of the process are calculated in a perturbative manner as sums of free moments.  

A derivation of the path integral representation can be found in \cite{GKO_PILNP_2017}. In case of the response function $h$ being chosen as $\alpha$\nobreakdash-function,
\cite{Campbell2014} derived an explicit path integral representation for the shot noise variable $s\fargs{t}=\sum_{t^\gamma} h\fargs{t-t^\gamma}$.

In brief, the line of reasoning in deriving the action of the LNP cascade model follows arguments from nonequilibrium statistical dynamics \cite{doi_second_1976,doi_stochastic_1976,peliti_pathintegral_1985,martinsiggiarose1973,tauber_field_2007,lefevre_biroli_2007}. 
For illustration, we consider the case $n=1$ and drop the neuron index $i$. Starting at time $t^0$ and discretizing time into $N_t$ steps of size $\dt$
\begin{equation}
 t^{\nu+1} = t^\nu +\dt,\quad \nu=0,1,\dots,N_t-1,
\end{equation}
we obtain a discrete spike count $N^\nu\mdef N\fargs{t^\nu}$ and a discretized intensity process $\lambda^\nu\mdef \lambda\fargs{t^\nu}$.
In the following, we set $N^0 = 0$.
Given conditionally independent Poissonian increments, the spike counting process reads
\begin{equation}\label{eq:discrete_model}
 N^{\nu+1}=N^\nu + \eta^\nu,\quad \eta^\nu\sim\Poiss\fargs{\dt\lambda^\nu}.
\end{equation}
Starting point in deriving the action is the pdf written in terms of $\delta$-functions constrained to the solutions of 
Eq.~\eqref{eq:discrete_model},
\begin{align}\nonumber\label{eq:Laplacerep_pdf}
 P\left[\left(N^1,\dots,N^{N_t}\right)|\left(\eta^0,\dots,\eta^{N_t-1}\right),N^0=0\right] = 
\prod_{\nu=0}^{N_t-1}\deltafct{N^{\nu+1}-N^\nu-\eta^\nu}\\
=\int \left(\prod_{\nu=0}^{N_t-1}\frac{\intd \tilde z^\nu}{2\pi\ci}\right)\exp\left(-\sum_\nu 
\tilde z^\nu\left(N^{\nu+1}-N^\nu-\eta^\nu\right)\right).
\end{align}
In the second step, the Laplace representation 
\begin{equation}
 \deltafct{x} = \frac{1}{2\pi\ci}\int e^{\tilde z x}\intd\tilde z 
\end{equation}
of the $\delta$-function has been used for all $\nu$.
Note that the integration over $\tilde z$ is along the imaginary axis.
 
In general, the Poisson probability density function for variables $k$ and mean parameter $\lambda$ is given by
\begin{equation}
 P_\lambda\fargs{k}=\frac{\lambda^k}{k!}e^{-\lambda}.
\end{equation}
The moment generating function for such a process reads
\begin{align}\nonumber
 Z\larg[J\rarg] = \ensexpect{e^{Jk}} &= \sum_{k=0}^\infty\frac{\lambda^k}{k!}e^{-\lambda}e^{Jk} = 
e^{-\lambda}e^{\lambda e^J} \\
  &= \exp\left(\lambda\left(e^J-1\right)\right).
\end{align}
Noise increments $\eta^\nu$ are conditionally independent, which implies 
\begin{equation}
 P\larg[\eta^0,\dots,\eta^{N_t-1}\rarg]=\prod_\nu P\fargs{\eta^\nu|\lambda^\nu}.
\end{equation}
Marginalizing the pdf \eqref{eq:Laplacerep_pdf} over the noise increments yields
\begin{align}\nonumber\label{eq:marginalized_pdf}
 P\larg[\left(N^1,\dots,N^{N_t}\right)|N^0=0\rarg] &= \sum_{\eta^0}\dots\sum_{\eta^{N_t-1}}\\ &\hspace*{-72bp}
P\larg[\left(N^1,\dots,N^{N_t}\right)|\left(\eta^0,\dots,\eta^{N_t-1}\right),N^0=0\rarg]
\prod_{\nu=0}^{N_t-1} 
P_{\dt\lambda^\nu}\fargs{\eta^\nu}.
\end{align}
Note that for one particular noise increment one has
\begin{align}\nonumber
\sum_{\eta^\nu}e^{-\tilde z^\nu\left(-\eta^\nu\right)}\cdot 
\frac{\left(\dt\lambda^\nu\right)^{\eta^\nu}}{\eta^\nu!}e^{-\dt \lambda^\nu} &= 
\ensexpect{e^{\tilde z^\nu\eta^\nu}} \\
&= Z\larg[\tilde z^\nu\rarg] = \exp\left(\dt\lambda^\nu\left(e^{\tilde z^\nu}-1\right)\right).
\end{align}
Substituting this result into the marginalized pdf \eqref{eq:marginalized_pdf} yields
\begin{align}\nonumber
 P\larg[\left(N^1,\dots,N^{N_t}\right)|N^0=0\rarg] = \int \left(\prod_{\nu=0}^{N_t-1}\frac{\intd 
\tilde z^\nu}{2\pi\ci}\right)& \exp\left(-\sum_\nu\tilde z^\nu\left(N^{\nu+1}-N^\nu\right)\right)\\
  & \times\exp\left(\sum_\nu \dt\lambda^\nu\left(e^{\tilde z^\nu}-1\right)\right).
\end{align}

As a next step, the limit of infinitesimal step size $\dt$ and infinitely many steps ($N_t\rightarrow\infty$) is performed, while 
keeping the total time $T=N_t\dt$ constant. This turns sums into integrals
\begin{equation}
 \lim_{\dt\rightarrow 0} \sum_{\nu=0}^{N_t=T/\dt} \dt\cdot f\fargs{\nu \dt} = \int_0^T f\fargs{t}\intd t .
\end{equation}
The full path integral then reads
\begin{align}\nonumber
 P\larg[z|z\fargs{0}=0\rarg] = \int \PId\tilde z & \exp\left\{-\int_{t^0}^T \tilde z\fargs{t} z\fargs{t}\intd t\right.\\
 & \qquad +\left.\int_{t^0}^T \lambda\fargs{t}\left(e^{\tilde z\fargs{t}}-1\right)\intd t\right\},
\end{align}
with $\PId\tilde z = \lim_{N_t\rightarrow \infty}\prod_{\nu=0}^{N_t-1}\frac{\intd\tilde z^\nu}{2\pi\ci}$ and
$\lim_{\dt\rightarrow 0} \frac{N\fargs{t+\dt}-N\fargs{t}}{\dt} = z\fargs{t}$.
Conventionally one writes the pdf $P\left[z|z\fargs{0}=0\right]$ as a path 
integral over the so-called path density $\mathfrak{p}\larg[\tilde z,z\rarg]$ which is itself an exponential of the negative action $-S\larg[\tilde z,z\rarg]$
\begin{align}\nonumber\label{eq:pdf}
 P\larg[z|z\fargs{0}=0\rarg] &= \int \PId\tilde z\exp\left(-S\larg[\tilde z,z\rarg]\right) \\ &= 
 \int \PId\tilde z~ \mathfrak{p}\larg[\tilde z,z\rarg] .
\end{align}
Thus, the action for a LNP cascade process for $n=1$ reads
\begin{equation}
 S\larg[\tilde z,z \rarg] = \int \tilde z\fargs{t} z\fargs{t} - \left(e^{\tilde z\fargs{t}}-1\right)\phi\fargs{\conv{h}{z}\fargs{t}+b\fargs{t}}\intd t .
\end{equation}

As everything factorizes, the action of the LNP model for more than one neuron ($n>1$) is finally given by \cite{GKO_PILNP_2017},
\begin{equation}
 S_\mathrm{LNP}\actarg=\sum_i \int \tilde z_i\fargs{t} z_i \fargs{t} - \left(e^{\tilde z_i \fargs{t}}-1\right)\phi\fargs{\sum_j\conv{h_{ij}}{z_j}\fargs{t}+b_i\fargs{t}}\intd t.
\end{equation}

Proceeding with the action $S_\mathrm{LNP}\actarg$ directly would yield an expansion about ${z = 0}$. 
However, a different expansion point can be chosen by shifting the configuration and response variables by $\bar r$ and $\tilde r$, respectively.
Transforming the variables according to
\begin{align}
 z_i &\longrightarrow \delta z_i \mdef z_i - \bar r_i, \\
 \tilde z_i &\longrightarrow \delta \tilde z_i \mdef \tilde z_i - \tilde r_i,
\end{align}
results in an action $S^{\ast}\larg[\delta\tilde z,\delta z\rarg]$.
Shifting the processes by its first moments, $\bar r_i=\ensexpect{z_i}$ and $\tilde r_i=\ensexpect{\tilde z_i}$, yields an \emph{effective action} \cite{Buice2007}.
As the action is at least linear in the auxiliary response variable, its first moment $\ensexpect{\tilde z_i}$ is zero and the transformation of $\tilde z_i$ is trivial. % (see also \cite{Stapmanns2018}).

The general Taylor expansion of $\phi$ about $\bar r$ is given by
\begin{align}
 \phi\fargs{\sum_j\conv{h_{ij}}{z_j}\fargs{t}+b_i\fargs{t}} &= \phi\fargs{\sum_j\conv{h_{ij}}{\delta z_j}\fargs{t}+\sum_j\conv{h_{ij}}{\bar r_j}\fargs{t}+b_i\fargs{t}}\\ 
&= \sum_{q=0}^\infty\frac{1}{q!}\phi^{(q)}_i \conv{h}{\delta z}_i^q\fargs{t},
 \end{align}
where $\conv{h}{\delta z}_i\fargs{t} = \sum_j \conv{h_{ij}}{\delta z_j}\fargs{t}$ and coefficients
\begin{equation}\label{eq:phi_coeff}
 \phi^{(q)}_i = \frac{\diffd^q}{\diffd x^q}\phi\fargs{x}\arrowvert_{\conv{h}{\bar r}_i+b_i}.
\end{equation}
Raising the convolution to the power of $q$ is meant to be
\begin{equation}
 \conv{h}{\delta  z}_i^q\fargs{t}=\underbrace{\conv{h}{\delta  z}_i\fargs{t}\times \dots \times \conv{h}{\delta  z}_i\fargs{t}}_{q~\text{times}}.
\end{equation}
The action with transformed variables then reads
\begin{align}\nonumber\label{eq:S_star}
 S^\ast\effarg &= \sum_i\int\intd t \Biggl[\tilde z_i\fargs{t}\delta z_i\fargs{t} + \tilde z_i\fargs{t}\bar r_i\fargs{t}\\
&  \qquad -\underbrace{\left(e^{\tilde z_i\fargs{t}}-1\right)}_{=\sum_{p=1}^\infty\frac{\left(\tilde z_i\fargs{t}\right)^p}{p!}}
\left(\sum_{q=0}^\infty\frac{1}{q!}\conv{h}{\delta z}_i^q\fargs{t}\right)\Biggr].
 \end{align}

For moment calculations, the bilinear part of the action is split off
\begin{equation}\label{eq:Sstar_splitting}
 S^\ast\effarg = S_F\effarg + S^\ast_I\effarg,
\end{equation}
where the free (bilinear) action reads 
\begin{equation}\label{eq:free_action}
 S_F\effarg = \sum_i\int\intd t \tilde z_i\fargs{t}\left(\delta z_i\fargs{t}-\phi_i^{(1)}\conv{h}{\delta z}_i\right).
\end{equation}
The component representing the interaction is given by
\begin{align}\label{eq:interacting_staraction}\nonumber
 S^\ast_I\effarg &= -\sum_i\int \intd t \sum_{\substack{p= 1,q = 0\\\setminus\set{p=q=1}\\\setminus\set{p=1,q=0}}} 
 \frac{\left(\tilde z_i\fargs{t}\right)^p}{p!}
 \frac{\phi_i^{(q)}}{q!}\conv{h}{\delta z}_i^q\fargs{t}\\
&\qquad +\sum_i\int\intd t \tilde z_i\fargs{t}\left(\bar r_i\fargs{t}-\phi_i^{(0)}\right).
\end{align}
The \emph{inverse tree-level propagator} is defined by the bilinear part $S_F$ of the action, 
\begin{equation}\label{eq:def_invtreelevelprop}
\invtreeprop{ij}\fargs{t,t^\prime}\mdef\delta_{ij}\deltafct{t-t^\prime} - \phi_i^{(1)} h_{ij}(t-t^\prime).
\end{equation}
Finally, the tree-level propagator can be deduced from 
\begin{equation}\label{eq:treeprop_cond}
 \sum_k \int\intd t^\prime \invtreeprop{ik}\fargs{t,t^\prime}\treeprop{kj}\fargs{t^\prime,t^\dprime} = \deltafct{t-t^\dprime}\delta_{ij}.
\end{equation}
In this representation, \emph{tree} and \emph{loop} level refer to zeroth order and first order, respectively, in a small-fluctuation expansion in terms of higher-order moments coupling to lower-order ones \cite{Buice2007,Chow2015}. 
This expansion is called semiclassical expansion, or loop expansion, because the number of loops in a diagram corresponds to higher-order fluctuation contributions \cite{Chow2015,GKO_PILNP_2017}. Whenever fluctuations are small, the truncated loop expansion provides a good approximation for spike train statistics.

In the following, the expansion is performed about the mean-field solution, instead of choosing $\bar r=0$.
Very recently, \cite{Stapmanns2018} used renormalization group techniques to self-consistently determine the statistics of the process.
An application of these methods to our model still needs to be investigated.
Here, we consider an approximation, where the expansion point is determined by the free (tree-level) expectation
\begin{equation}\label{eq:SMF_expansionpoint}
 \bar r_i\fargs{t} = \freeexpect{z_i\fargs{t}}=\phi^{(0)}_i,
\end{equation}
where $\freeexpect{\cdot}$ denotes the expectation with respect to the free path density $\exp\fargs{-S_F}$.
The true mean of the process is different for nonlinear gain functions, as there are non-vanishing loop corrections to the mean (cf.\ Section~\ref{sec:meancorr}).

By Eq.~\eqref{eq:SMF_expansionpoint}, the last term in Eq.~\eqref{eq:interacting_staraction} vanishes and the interacting action reads
\begin{equation}
  S^\mathrm{MF}_I\effarg = -\sum_i\int \intd t \sum_{\substack{p= 1,q = 0\\\setminus\set{p=q=1}\\\setminus\set{p=1,q=0}}} 
 \frac{\left(\tilde z_i\fargs{t}\right)^p}{p!}
 \frac{\phi_i^{(q)}}{q!}\conv{h}{\delta z}_i^q\fargs{t}.
\end{equation}
In total, the action of the LNP model expanded about its mean-field solution reads
\begin{equation}\label{eq:SMF_splitting}
 S^\mathrm{MF}\effarg = S_F\effarg + S^\mathrm{MF}_I\effarg.
\end{equation}
The path-integral representation of the probability density functional is given by
\begin{equation} 
 P\left[\delta z|\delta z\fargs{0}\right] = \int \PId\tilde z(t) \exp\fargs{-S^\mathrm{MF}\effarg}, 
\end{equation}
where $\PId \tilde z$ denotes the path integral measure. This implies a joint moment-generating functional of $\delta z$ and $\tilde z$
\begin{equation}
 Z\larg[J,\tilde J\rarg] = \int\!\PId z\fargs{t}\int\!\PId\tilde z\fargs{t} e^{-S^\mathrm{MF}\effarg +J\tilde z + \tilde J \delta z},
\end{equation}
and a moment-generating functional of $\delta z$
\begin{equation}
 Z\larg[\tilde J\rarg] = \int\!\PId z\fargs{t} e^{-S^\mathrm{MF}\effarg + \tilde J \delta z}.
\end{equation}

As a consequence of the splitting in Eq.~\eqref{eq:SMF_splitting}, arbitrary moments of the process can be calculated as a combination of moments of the free action
\begin{align}\label{eq:eff_diagram_expansion}\nonumber
 \ensexpect{\prod_{\iota=1}^l \delta z_{i_\iota}\fargs{t_\iota} \prod_{\varsigma=1}^m \tilde z_{i_\varsigma}\fargs{t_\varsigma}} &= \int\PId\delta z\PId\tilde z 
 \prod_{\iota=1}^l \delta z_{i_\iota}\fargs{t_\iota} \prod_{\varsigma=1}^m \tilde z_{i_\varsigma}\fargs{t_\varsigma}
 e^{-S^\mathrm{MF}\effarg} \\
 &\hspace*{-82bp}=\freeexpect{\prod_{\iota=1}^l \delta z_{i_\iota}\fargs{t_\iota} \prod_{\varsigma=1}^m \tilde z_{i_\varsigma}\fargs{t_\varsigma}
 \smash{\prod_{\substack{p,q = 1\\\setminus\set{p=q=1}\\\setminus\set{p=1,q=0}}}}\sum_{k=0}^\infty\frac{1}{k!}\left(\frac{\tilde z_i^p}{p!}
 \frac{\phi_i^{(q)}}{q!}\conv{h}{\delta z}_i^q\right)^k}, %\\[-20bp]\nonumber
\end{align}
where $\freeexpect{\cdot}$ again denotes the expectation with respect to the free path density $\exp\fargs{-S_F}$.
By completing the square and using the definition of the inverse tree-level propagator, the moment-generating functional of the free moments reads
\begin{equation}
 Z_F\left[\tilde J,J\right] = \exp\left\{\int\intd t \int\intd t^\prime \tilde 
J(t)\treeprop{}\fargs{t,t^\prime}J(t^\prime)\right\}.
\end{equation}
This functional is bilinear and thus only products of tree-level propagators survive when calculating free moments
\begin{equation}\label{eq:freemoment_expansion}
\freeexpect{\prod_{\iota=1}^l \delta z_{i_\iota}\fargs{t_\iota} \prod_{\varsigma=1}^m \tilde z_{i_\varsigma}\fargs{t_\varsigma}} 
%  \ensexpect{\tilde z^l\delta z^m}
 = \sum_{\substack{\text{pair-wise}\\ \text{partitions}}}\quad\prod_{\mathrm{pairs}~(\iota,\varsigma)} \treeprop{i_\iota i_\varsigma}\fargs{t_\iota,t_\varsigma}.
\end{equation}

The expansion of Eq.~\eqref{eq:eff_diagram_expansion} can be represented diagrammatically by use of the observation~\eqref{eq:freemoment_expansion}. Contributions from individual components of the series are obtained by the Feynman rules listed in the following section and derived in \cite{GKO_PILNP_2017}.

\subsubsection{\label{sec:FeynmanRules}Feynman rules of the LNP cascade model}
The Feynman rules in this section are derived for computing cumulant expansions about the mean-field solution $\bar r=\freeexpect{z}$,
considering $S^\mathrm{MF}$ from Eq.~\eqref{eq:SMF_splitting}. 
% Note that due to self-consistency 
% $\mu_i\fargs{t}\mdef \bar r_i\fargs{t} - \phi^{(0)}_i = 0$ in Eq.~\eqref{eq:interacting_action}.
For bookkeeping reasons, each term in Eq.~\eqref{eq:eff_diagram_expansion} is represented by a Feynman diagram.
The building blocks of these diagrams are
\begin{itemize}
 \item external vertices according to the desired moment
	\begin{equation*}
	\begin{picture}(70,10)(0,0)%\input{./../figures/tikz/FD_ext_vert_z.tikz}
	    \begin{tikzpicture}[scale=1.,baseline=-.5ex,node distance=.8cm and 0.8cm]
        % First we draw the vertices
        \node[outervertexstyle] (cl) {};
        \node[right=of cl] (b) {};
        \node[right=of b] (c) {};
        % Connect vertices with edges and draw weights
        \path[arrowedge] (b) to (cl);
        \path[dottededge] (c) to (b);
        \end{tikzpicture}
	\end{picture} = 1
	\end{equation*}
	\begin{equation*}
	\begin{picture}(70,10)(0,0)%\input{./../figures/tikz/FD_ext_vert_tildez.tikz}
	    \begin{tikzpicture}[scale=1.,baseline=-.5ex,node distance=.8cm and 0.8cm]
        % First we draw the vertices
        \node (cl) {};
        \node[right=of cl] (b) {};
        \node[outervertexstyle, right=of b] (c) {};
        % Connect vertices with edges and draw weights
        \path[dottededge] (b) to (cl);
        \path[arrowedge] (c) to (b);
        \end{tikzpicture}
	\end{picture} = 1
	\end{equation*}

 \item the propagator
	\begin{equation*}
	\begin{picture}(80,20)(0,0)%\input{./../figures/tikz/FD_prop.tikz}
	    \begin{tikzpicture}[scale=1.,baseline=-.5ex,node distance=.8cm and 0.8cm]
        % First we draw the vertices
        \node[label=above:$t_1$,label=below:{\scriptsize $i$}] (cl) {};
        \node[right=1.6cm of cl,label=above:$t_2$,label=below:{\scriptsize $j$}] (b) {};
        % Connect vertices with edges and draw weights
        \path[arrowedge] (b) to (cl);
        \end{tikzpicture}
	\end{picture} = \treeprop{ij}\fargs{t_1,t_2}
	\end{equation*}
 \item the filter edge
	\begin{equation*}
	\begin{picture}(80,20)(0,0)%\input{./../figures/tikz/FD_filteredge.tikz}
        \begin{tikzpicture}[scale=1.,baseline=-.5ex,node distance=.8cm and 0.8cm]
        % First we draw the vertices
        \node[label=above:$t_2$,label=below:{\scriptsize $i$}] (cl) {};
        \node[right=1.6cm of cl,label=above:$t_1$,label=below:{\scriptsize $j$}] (b) {};
        % Connect vertices with edges and draw weights
        \path[filteredge] (b) to (cl);
        \end{tikzpicture}
	\end{picture} = h_{ij}\fargs{t_2-t_1}
	\end{equation*}
 \item the filter vertex
	\begin{equation*}
	\begin{picture}(80,20)(0,0)%\input{./../figures/tikz/FD_filtervertex.tikz}
        \begin{tikzpicture}[scale=1.,baseline=-.5ex,node distance=.8cm and 0.8cm]
        % draw vertices
        \node[label=below:{\scriptsize $i$}] (cl) {};
        \node[innervertexstyle,right=of cl,label=above:$t$] (m) {};
        \node[right=of m,label=below:{\scriptsize $j$}] (cr) {};
        % % Connect vertices with edges and draw weights
        \foreach \source/ \dest in {cr/m}
            \draw[arrowedge] (\source) -- (\dest);
        \foreach \source/ \dest in {m/cl}
            \draw[filteredge] (\source) -- (\dest);
        \end{tikzpicture}
	\end{picture} = \delta_{ij}
	\end{equation*}
 \item the internal vertices
	\begin{align*}
	\begin{picture}(120,40)(0,0)%\input{./../figures/tikz/FD_genvertex.tikz}
        \begin{tikzpicture}[scale=1.,baseline=-.5ex,node distance=.8cm and 0.8cm]
        % draw vertices
        \node (cl) {};
        \node[innervertexstyle,right=of cl,label=above:$t$] (m) {};
        \node[below right=of m,label={[label distance=-.4cm]-45:{\scriptsize $j_q$}}] (k1) {};
        \node[above right=of m,label={[label distance=-.4cm]45:{\scriptsize $j_1$}}] (kq) {};
        \node[above left=of m,label={[label distance=-.4cm]above left:{\scriptsize $i_1$}}] (p1) {};
        \node[below left=of m,label={[label distance=-.4cm]below left:{\scriptsize $i_p$}}] (pp) {};
        % % Connect vertices with edges and draw weights
        \foreach \source/ \dest in {m/p1,m/pp}
            \draw[arrowedge] (\source) -- (\dest);
        \foreach \source/ \dest in {k1/m,kq/m}
            \draw[filteredge] (\source) -- (\dest);
        \path[dottededge] (1.6,.3) to node[right=2pt]{$q$ legs} (1.6,-.3);
        \path[dottededge] (.4,.3) to node[left=2pt]{$p$ legs}(.4,-.3);
        \end{tikzpicture}
	\end{picture} &= \frac{\phi_{i_1}^{(q)}}{p!q!}\prod_{\iota=2}^p\delta_{i_1i_\iota}\prod_{\varsigma=1}^q\delta_{i_1j_\varsigma},\\\nonumber 
	\end{align*}
	where $\phi_i^{(q)}$ has been defined in Eq.~\eqref{eq:phi_coeff}. From Eq.~\eqref{eq:eff_diagram_expansion} it is obvious that $p\geq 2$ for $q=0$. Otherwise, any combination is allowed such that $q+p\geq 3$.
\end{itemize}
The moment expansion given in Eq.~\eqref{eq:eff_diagram_expansion} allows to deduce an algorithmic recipe to calculate cumulants. Note that these are calculated about the mean-field solution $\bar r$, i.e.\ $\treeexpect{\delta z}=0$.
The $n^\mathrm{th}$ order cumulant is obtained by the following recipe
\begin{enumerate}
 \item Determine all possible graphs, where the cumulant order corresponds to the external vertices and the perturbation order corresponds to inner vertices.
 \item Translate vertices and edges into formulae according to the building blocks described above.
 \item Integrate over internal vertex times $\int\intd t_\xi$ and sum over neuron indices.
 \item If a vertex type occurs $k$ times (from $\left(S_I\effarg\right)^k$ terms) include a factor $1/k!$.
 \item For $n_\mathrm{sym}$ topologically identical graphs, multiply with the factor $n_\mathrm{sym}$ (corresponding to possible pairings in Eq.~\eqref{eq:freemoment_expansion}).
\end{enumerate}
A frequently recurring sub-diagram is the combination of a filter edge, a filter vertex, and the propagator. We therefore introduce the shorthand
\begin{equation}\label{eq:loopedge}
\convfirst{h}{\treeprop{}}_{ij}\fargs{t,t^\dprime} \mdef \sum_k\int \intd t^\prime h_{ik}\fargs{t-t^\prime}\treeprop{kj}\fargs{t^\prime,t^\dprime}.
\end{equation}
We will use these rules to deduce explicit predictions in the following section.

%%%%%%%%%%%%%%%%%
\section{Results}
%%%%%%%%%%%%%%%%%
%%%%%%%%%%%%%%%%%

\subsection{Techniques for calculating loop corrections}
A procedure to compute higher-order cumulants and their corrections arising from the nonlinear transfer function was described in the previous section.% \eqref{eq:intensity_LNP}.
However, there are some technical obstacles to actually calculate them. 
For an example, consider the one-loop correction to the firing rates
\begin{align}\nonumber
\ensexpect{\delta z_i}_\mathrm{1-loop} &=
\begin{picture}(70,40)(0,0)%\input{./../figures/tikz/FD_rate1loop.tikz}
    \begin{tikzpicture}[scale=1.,baseline=-.5ex,node distance=.8cm and 0.8cm]
    % draw vertices
    \node[outervertexstyle,label=above:$t$,label=below:{\scriptsize $i$}] (cl) {};
    \node[innervertexstyle,right=of cl,label=above:$t_1~$,label=below:{\scriptsize $j$}] (l4) {};
    \node[innervertexstyle,below right=of l4] (l1) {};
    \node[innervertexstyle,above right=of l4] (l3) {};
    \node[innervertexstyle,above right=of l1,label=above:$t_2$,label=below:{\scriptsize $l$}] (l2) {};
    % % Connect vertices with edges and draw weights
    \foreach \source/ \dest in {l4/cl,l2/l3,l2/l1}
        \draw[arrowedge] (\source) -- (\dest);
    \foreach \source/ \dest in {l3/l4,l1/l4}
        \draw[filteredge] (\source) -- (\dest);
    \end{tikzpicture}
\end{picture}\\\nonumber
% \vspace{20bp}
\\\nonumber
&= 2\cdot \sum_{j,l} \int \treeprop{ij}\fargs{t,t_1} \frac{\phi_j^{(2)}}{1!2!} 
\convfirst{h}{\treeprop{}}_{jl}\fargs{t_1,t_2} \convfirst{h}{\treeprop{}}_{jl}\fargs{t_1,t_2}\\\nonumber
&\hspace*{72bp} \times\frac{\phi_l^{(0)}}{2!} \intd t_1 \intd t_2. % \\\nonumber
% &= 2\cdot\sum_{j,l}\int \FTtreeprop{\omega_1}_{ij}\frac{\phi_j^{(2)}}{1!2!}\frac{1}{2\pi}\deltafct{\omega_2+\omega_3-\omega_1}\\
% &\qquad \times \sum_{k_1} \FT{h}_{jk_1}\larg(\omega_2\rarg)\FTtreeprop{\omega_2}_{k_1l}
% 	\sum_{k_2} \FT{h}_{jk_2}\larg(\omega_3\rarg)\FTtreeprop{\omega_3}_{k_2l} \frac{\phi_l^{(0)}}{2!}\deltafct{-\omega_2-\omega_3}\intd \omega_1 \intd \omega_2 \intd \omega_3 \\.
\end{align}
The depicted Feynman diagram is the only one possible with one loop for the first-order cumulant.
It is translated into a formula in accordance with the Feynman rules described in Section~\ref{sec:FeynmanRules}. 
For a stationary process, the one-loop correction of the mean is time-independent. Computations are simpler in the Fourier domain, as temporal integration is eliminated
\begin{align}\nonumber
\ensexpect{\delta z_i}_\mathrm{1-loop} &= \frac{1}{4\pi}\sum_{j,k_1,k_2,l}\int \FTtreeprop{ij}\fargs{0}\phi_j^{(2)} \FT{h}_{jk_1}\larg(-\omega\rarg)\FTtreeprop{k_1l}\fargs{-\omega} \\
&\hspace*{72bp} \times \FT{h}_{jk_2}\fargs{\omega}\FTtreeprop{k_2l}\fargs{\omega} \phi_l^{(0)}\intd \omega .
\end{align}
Further details on this calculation can be found later in Section~\ref{sec:meancorr}.
Here we just note that one is left with the frequency integration
\begin{equation}
\sum_{k_1,k_2}\int_{-\infty}^{\infty} \FT{h}_{jk_1}\larg(-\omega\rarg)\FTtreeprop{k_1l}\fargs{-\omega} \FT{h}_{jk_2}\fargs{\omega}\FTtreeprop{k_2l}\fargs{\omega} \intd \omega.
\end{equation}
The authors in \cite{GKO_PILNP_2017} solved these integrals numerically by performing Riemann summation for a fixed range of frequencies.
Due to the analytic nature of the integrands, however, this approach is error-prone and may even fail completely.

In the following, a technique is presented which allows to calculate these integrals analytically for common choices of the response functions. 
After formalizing the loop integrals, all results will be assembled. As a consequence, the following section is rather technical.
We are using the explicit computations of first-order and second-order cumulants in Section~\ref{sec:explicit_preds} and compare them to a numerical simulation in Section~\ref{sec:numres_structLNP}.

\subsection{\label{sec:genloopint}General loop integrals}
A frequently occurring one-loop integral is characterized by the number of propagators in the loop
\begin{center}
% general loop diagram
%\input{./../figures/tikz/genloop.tikz}
    \begin{tikzpicture}[scale=1.,baseline=-.5ex,node distance=1.cm and 0.8cm]
    % draw vertices
    \node[innervertexstyle,label=right:{\tiny $~l_N=k_M$}] (cl) {};
    \node[innervertexstyle,below right=of cl,label=below left:$\nu_{N-1}+\omega$] (nn3c) {};
    \node[innervertexstyle,below right=of nn3c,label=above right:{\tiny $\!l_{N-1}$}] (nn3) {};
    \node[innervertexstyle,above right=of cl,label=above left:$\omega_{M-1}-\omega$] (wn3c) {};
    \node[innervertexstyle,above right=of wn3c,label=below right:{\tiny $\!k_{M-1}$}] (wn3) {};
    \node[right=.3cm of wn3] (udots) {$\cdots$};
    \node[right=.3cm of nn3] (ldots) {$\cdots$};
    \node[innervertexstyle,right=.3cm of udots,label=below:{\tiny $k_{2}$}] (wn2) {};
    \node[innervertexstyle,right=.3cm of ldots,label=above:{\tiny $l_{2}$}] (nn2) {};
    \node[innervertexstyle,right=of wn2,label=above:$\omega_1-\omega$] (wn2c) {};
    \node[innervertexstyle,right=of nn2,label=below:$\nu_1+\omega$] (nn2c) {};
    \node[innervertexstyle,right=of wn2c,label=below:{\tiny $k_{1}~$}] (wn1) {};
    \node[innervertexstyle,right=of nn2c,label=above:{\tiny $l_{1}$}] (nn1) {};
    \node[innervertexstyle,below right=of wn1,,label=above right:$\omega_0-\omega$] (wn1c) {};
    \node[innervertexstyle,below right=of wn1c,label=left:{\tiny $l_{0}=k_0$}] (cr) {};
    \node[innervertexstyle,above right=of nn1,label=below right:$\omega$] (nn1c) {};
    % % Connect vertices with edges and draw weights
    \foreach \source/ \dest in {cr/nn1c,cr/wn1c,wn1/wn2c,wn3/wn3c,nn1/nn2c,nn3/nn3c}
        \draw[arrowedge] (\source) -- (\dest);
    \foreach \source/ \dest in {wn1c/wn1,wn2c/wn2,wn3c/cl,nn1c/nn1,nn2c/nn2,nn3c/cl}
        \draw[filteredge] (\source) -- (\dest);
    \end{tikzpicture}
\end{center}
Note that legs entering or leaving the loop have been omitted at this point, and an ambiguity in the indices ($l_0=k_0$ and $l_N=k_M$) is introduced for ease of notation. Later, this issue will be resolved by appropriate Kronecker $\delta$'s.
To calculate corrections involving these loops as part of the diagram, we define a general loop integral in Fourier space
\begin{align}\label{eq:Lloop_def}\nonumber
 \mathcal{L}_{\Lloopidx}&\fargs{\vec{\omega},\vec{\nu}} = \int E_{k_Mk_{M-1}}\fargs{\omega_{M-1}-\omega}\cdots E_{k_2k_1}\fargs{\omega_1-\omega}\\
 &\hspace*{-30bp} \times E_{k_1k_0}\fargs{\omega_0-\omega} E_{l_Nl_{N-1}}\fargs{\nu_{N-1}+\omega}\cdots E_{l_2l_1}\fargs{\nu_1+\omega}E_{l_1l_0}\fargs{\omega} \intd \omega
\end{align}
where $\vec\omega = \left(\omega_0,\dots,\omega_{M-1}\right)$ and $\vec\nu=\left(\nu_1,\dots,\nu_{N-1}\right)$.
A loop edge $E$ in Fourier space is the combination of a filter edge and a propagator (cf.\ Eq.~\eqref{eq:loopedge}) and is given by
\begin{equation}
 E_{ij}\fargs{\omega} = \sum_k \FT{h}_{ik}\fargs{\omega}\FTtreeprop{kj}\fargs{\omega}.
\end{equation}

For an analytic solution, we aim to calculate the integrals $\mathcal{L}$. First, note that the $\omega$\nobreakdash-dependence of the Fourier transformed propagator $\FTtreeprop{}\fargs{\omega}$ is completely determined by the Fourier transform $\FT{h}\fargs{\omega}$ of the response function, which is assumed to be the same for all neurons. We define $D_{\phi^{(1)}}\mdef \diagm{\phi_1^{(1)},\dots,\phi_n^{(1)}}$ and exploit a diagonalization of the matrix $D_{\phi^{(1)}}W$
\begin{equation}
 V^{-1}D_{\phi^{(1)}}WV=\diagm{\xi_{1},\dots,\xi_{n}},
\end{equation}
where $\xi_{i}$ denotes the $i$\nobreakdash-th eigenvalue of $D_{\phi^{(1)}}W$. 
Assuming that the matrix $D_{\phi^{(1)}}W$ is diagonalizable considerably simplifies all subsequent calculations.
If this is not the case, one would consider the Jordan canonical form and extend the following computations in a suitable way.
For the Fourier transformed tree-level propagator determined from Eq.~\eqref{eq:treeprop_cond}, it follows that
\begin{align}\nonumber
\FTtreeprop{}\fargs{\omega} &= \left(\one - D_{\phi^{(1)}}\FT{h}\fargs{\omega}W \right)^{-1} \\
  & = V\diagm{\frac{1}{1-\FT{h}\fargs{\omega}\xi_{1}},\dots,\frac{1}{1-\FT{h}\fargs{\omega}\xi_{n}}} V^{-1} .
\end{align}
Defining
\begin{equation}\label{eq:Ddef}
 D_i\fargs{\omega} = \frac{\FT{h}\fargs{\omega}}{1-\FT{h}\fargs{\omega}\xi_{i}}
\end{equation}
and assuming the same response function $h$ for all neurons, the loop edge $E$ reads
\begin{equation}
 E_{ij}\fargs{\omega} = \sum_k \left(WV\right)_{ik} D_k\fargs{\omega} V^{-1}_{kj}.
\end{equation}
The integrals to be solved are
\begin{align}\label{eq:Iloop_int}
 \mathcal{I}_\Iloopidx\fargs{\vec\omega,\vec\nu} = \int \prod_{\iota=0}^{M-1} D_{i_\iota}\fargs{\omega_\iota-\omega}
 D_{j_0}\fargs{\omega}\prod_{\varsigma=1}^{N-1}D_{j_\varsigma}\fargs{\nu_\varsigma+\omega}\intd\omega.
\end{align}
In what follows, we are sometimes only considering the integrand of these integrals $\mathcal{I}_\Iloopidx$ which is then denoted by $I_\Iloopidx$.
Given these integrals, the loop integral reads
\begin{align}\nonumber
 \mathcal{L}_{\Lloopidx}\fargs{\vec\omega,\vec\nu} &= \sum_{i_0,\dots i_{M-1}}\sum_{j_0,\dots,j_{N-1}} \mathcal{I}_\Iloopidx\fargs{\vec\omega,\vec\nu} \\
 &\qquad\times\prod_{\iota=0}^{M-1} \left(WV\right)_{k_{\iota+1}i_\iota}V_{i_\iota k_\iota}^{-1} \prod_{\varsigma=0}^{N-1} \left(WV\right)_{l_{\varsigma+1}j_\varsigma}V_{j_\varsigma l_\varsigma}^{-1} .
\end{align}

A solution of the integral in Eq.~\eqref{eq:Iloop_int} can be obtained using Cauchy's residue calculus \cite[Chapter~6]{lang_complexanalysis_1999}. % \cite[Chapter~13 \& 14]{remmert_funktionentheorie_2002}.
Let $\mathbb{H}=\set{z\in\mathbb{C}|\myIm\fargs{z}>0}$ denote the upper half-plane of the complex plane and $\overline{\mathbb{H}}=\mathbb{H}\cup\mathbb{R}$ its closure.
We consider a function $I$, which is holomorphic in $\overline{\mathbb{H}}$ apart from finitely many points which do not lie on the real axis. We further assume that the integral $\mathcal{I}=\int_{-\infty}^\infty I\fargs{x}\intd x$ exists and $\lim_{z\to\infty}zI\fargs{z}=0$. Then Cauchy's contour can be closed in the upper half-plane $\mathbb{H}=\set{z\in\mathbb{C}|\myIm\fargs{z}>0}$.
\begin{center}
    \begin{tikzpicture}[scale=.5]
    \draw (0,-.5) -- (0,6.) node[above] {$\myIm(z)$};  % Axis
    \draw (-5.5,0) -- (5.5,0) node[right] {$\myRe(z)$};   
    \draw (-4pt,5) -- (4pt,5) node[pos=0,above left] {$r$};
    \draw (-5,-4pt) -- (-5,4pt) node[pos=0,below] {$-r$};
    \draw (5,-4pt) -- (5,4pt) node[pos=0,below] {$r$};
    \node at (-2,2) {$\times$}; 
    \node at (-2.5,1.5) {$a$}; 
    \draw[very thick,petrol,postaction={decorate,decoration={
        markings,
        mark=at position 0.2 with {\arrow[scale=1]{>}}}}]
     (5,0) arc (0:180:5);
    \draw[very thick,petrol,postaction={decorate,decoration={
        markings,
        mark=at position 0.8 with {\arrow[scale=1]{>}}}}]
     (-5,0) -- (5,0);
    \node[petrol,very thick] at (2.5,3) {$\Gamma\fargs{r}$};
    \end{tikzpicture}
\end{center}
As a consequence of Cauchy's residue theorem, the improper integral can be written as \cite{lang_complexanalysis_1999}% \cite[Theorem~14.1.3]{remmert_funktionentheorie_2002},
\begin{equation}\label{eq:residuetheorem_I}
 \mathcal{I}=2\pi\ci\sum_{a\in\mathbb{H}}\ind_\Gamma\fargs{a}\residue{a}{I},
\end{equation}
where $\ind_\Gamma\fargs{a}$ is the winding number, which equals one for all poles and our specifically chosen integration contour $\Gamma\fargs{r}$.
For a proof, the limit $r\to\infty$ is taken and the complex periphery part is assessed by means of $\lim_{z\to\infty}zI\fargs{z}=0$.
Further note that, for poles of order one, the residues can be calculated by \cite{lang_complexanalysis_1999}%\cite[Theorem~13.1.3]{remmert_funktionentheorie_2002},
\begin{equation}\label{eq:orderone_residues}
 \residue{a}{I} = \lim_{\omega\to a} \left(\omega-a\right) I\fargs{\omega}.
\end{equation}
In the following two sections, we consider two popular response functions in neuroscience. Explicit results for loops with two or three edges are given.

%%%%%%%%%%%%%%%%%%%%%%%%%%%%%%%%%%%%%%%%%%%%%%%%%%%%%%%%%%%%%%
\subsubsection{\label{sec:example_expdecay}Example I: exponential-decay response function}
%%%%%%%%%%%%%%%%%%%%%%%%%%%%%%%%%%%%%%%%%%%%%%%%%%%%%%%%%%%%%%

For an exponentially decaying response function $h$
\begin{equation}
h\fargs{t} =\frac{1}{\tau} e^{-t/\tau} \thetafct{t},
\end{equation}
the Fourier transform is
\begin{equation}
 \FT{h}\fargs{\omega}=\frac{1}{\left(1+\ci\tau\omega\right)}.
\end{equation}
Thus, the elementary components $D_i$ in Eq.~\eqref{eq:Ddef} of the integrand $I$ of $\mathcal{I}$ from Eq.~\eqref{eq:Iloop_int} are given by
\begin{equation}
 \De_i\fargs{\omega}=\frac{1}{1+\ci\tau\omega-\xi_i}.
\end{equation}
Explicit calculus yields the following expressions for $k$\nobreakdash-point integrals. This terminology specifies the number of loop edges to be $k$.

\paragraph{Two-point integral}
The two-point integral with two loop edges reads
\begin{equation}
%  \mathcal{I}_{i,j}^e\fargs{\omega_0}=\int_{-\infty}^\infty\frac{\intd \omega}{\left(1+\ci\tau\left(\omega_0-\omega\right)-\xi_i\right)
%  \left(1+\ci\tau\omega-\xi_j\right)} 
 \mathcal{I}_{i,j}^e\fargs{\omega_0}=\int_{-\infty}^\infty \De_i\fargs{\omega_0-\omega}\De_j\fargs{\omega}\intd \omega ,
\end{equation}
and its integrand $I_{i,j}^e$ has (up to) two poles in $\mathbb{H}$
\begin{align}
 a^{(i)} &= \frac{\ci}{\tau}\left(\xi_i-1\right)+\omega_0, \\ 
 a^{(j)} &= -\frac{\ci}{\tau}\left(\xi_j-1\right)  .
\end{align}
The residues obtained from Eq.~\eqref{eq:orderone_residues} are 
\begin{align}
 \residue{a^{(i)}}{I_{i,j}^e} &= \frac{\ci}{\tau}\De_j\fargs{a^{(i)}} ,\\
 \residue{a^{(j)}}{I_{i,j}^e} &= -\frac{\ci}{\tau}\De_i\fargs{\omega_0-a^{(j)}}.
\end{align}
The integral $\mathcal{I}_{i,j}^e\fargs{\omega_0}$ is then given by Eq.~\eqref{eq:residuetheorem_I}. Assuming $\myRe{\xi_i}<1$ for all $i\in\set{1,\dots,n}$, only the pole $ a^{(j)}$ lies in the upper half-plane and the integral explicitly reads
\begin{equation}\label{eq:Iij_exp}
 \mathcal{I}_{i,j}^e\fargs{\omega_0}= \frac{2\pi}{\tau}\left(2-\xi_i-\xi_j+\ci\tau\omega_0\right)^{-1}.
\end{equation}
The assumption $\myRe{\xi_i}<1$ for all $i\in\set{1,\dots,n}$ is made for simplicity. If it does not hold, the poles inside the contour might be different, and the sum over residues in Eq.~\eqref{eq:residuetheorem_I} has to be evaluated accordingly. In the case that poles are on the real axis, as it is the case if $\myRe{\xi_i}=1$,
perturbing the poles by $\pm\ci\varepsilon$ for some small $\varepsilon>0$, performing the integration, and then taking the limit $\varepsilon\to 0$, yields the desired result.

\paragraph{Three-point integral}
If the loop has three edges, there are two integrals to be solved.
The integrand of the first integral,
\begin{equation}
%  \mathcal{I}_{i,jk}^e\fargs{\omega_0,\nu_1}=\int_{-\infty}^\infty\frac{\intd \omega}{\left(1+\ci\tau\left(\omega_0-\omega\right)-\xi_i\right)
%  \left(1+\ci\tau\omega-\xi_j\right)\left(1+\ci\tau\left(\nu_1+\omega\right)-\xi_k\right)} 
 \mathcal{I}_{i,jk}^e\fargs{\omega_0,\nu_1}=\int_{-\infty}^\infty\De_i\fargs{\omega_0-\omega}\De_j\fargs{\omega}\De_k\fargs{\nu_1+\omega}\intd \omega ,
\end{equation}
has poles 
\begin{align}
 a^{(i)} &= \frac{\ci}{\tau}\left(\xi_i-1\right)+\omega_0, \\ 
 a^{(j)} &= -\frac{\ci}{\tau}\left(\xi_j-1\right) ,\\ 
 a^{(k)} &= -\frac{\ci}{\tau}\left(\xi_k-1\right)-\nu_1.  
\end{align}
The residues in these poles are
\begin{align}
 \residue{a^{(i)}}{I_{i,jk}^e} &= \frac{\ci}{\tau}\De_j\fargs{a^{(i)}}\De_k\fargs{\nu_1+a^{(i)}}, \\
 \residue{a^{(j)}}{I_{i,jk}^e} &= -\frac{\ci}{\tau}\De_i\fargs{\omega_0-a^{(j)}}\De_k\fargs{\nu_1+a^{(j)}} , \\
 \residue{a^{(k)}}{I_{i,jk}^e} &= -\frac{\ci}{\tau}\De_i\fargs{\omega_0-a^{(k)}}\De_j\fargs{a^{(k)}} .
\end{align}
The other possible three-point integral,
\begin{equation}
%  \mathcal{I}_{ij,k}^e\fargs{\omega_0,\omega_1}=\int_{-\infty}^\infty\frac{\intd \omega}{\left(1+\ci\tau\left(\omega_0-\omega\right)-\xi_i\right)
%  \left(1+\ci\tau\left(\omega_1-\omega\right)-\xi_j\right)\left(1+\ci\tau\omega-\xi_k\right)} 
 \mathcal{I}_{ij,k}^e\fargs{\omega_0,\omega_1}=\int_{-\infty}^\infty 
 \De_i\fargs{\omega_0-\omega}
 \De_j\fargs{\omega_1-\omega}
 \De_k\fargs{\omega} \intd \omega
\end{equation}
has an integrand $I_{ij,k}^e$ with poles
\begin{align}
 a^{(i)} &=  \frac{\ci}{\tau}\left(\xi_i-1\right)+\omega_0, \\ 
 a^{(j)} &=  \frac{\ci}{\tau}\left(\xi_j-1\right)+\omega_1, \\ 
 a^{(k)} &= -\frac{\ci}{\tau}\left(\xi_k-1\right) ,
\end{align}
and corresponding residues
\begin{align}
 \residue{a^{(i)}}{I_{ij,k}^e} &=  \frac{\ci}{\tau}\De_j\fargs{\omega_1-a^{(i)}}\De_k\fargs{a^{(i)}}, \\
 \residue{a^{(j)}}{I_{ij,k}^e} &=  \frac{\ci}{\tau}\De_i\fargs{\omega_0-a^{(j)}}\De_k\fargs{a^{(j)}},\\
 \residue{a^{(k)}}{I_{ij,k}^e} &= -\frac{\ci}{\tau}\De_i\fargs{\omega_0-a^{(k)}}\De_j\fargs{\omega_1-a^{(k)}} .
\end{align}
The analytic form of the integrals $\mathcal{I}_{i,jk}^e\fargs{\omega_0,\nu_1}$ and $\mathcal{I}_{ij,k}^e\fargs{\omega_0,\omega_1}$ is again given by Eq.~\eqref{eq:residuetheorem_I} and greatly simplifies when assuming $\myRe{\xi_i}<1$ for all $i\in\set{1,\dots,n}$.

%%%%%%%%%%%%%%%%%%%%%%%%%%%%%%%%%%%%%%%%%%%%%%%%%%%%%%%%%%%%%%%%%%%%%%%%%%%%%%%%%%%
\subsubsection{Example II: \texorpdfstring{$\alpha$}{alpha}-type response function}
%%%%%%%%%%%%%%%%%%%%%%%%%%%%%%%%%%%%%%%%%%%%%%%%%%%%%%%%%%%%%%%%%%%%%%%%%%%%%%%%%%%

Another commonly used response function $h$ is given by
\begin{equation}\label{eq:alphafunc_def}
 h\fargs{t} =\frac{t}{\tau^2} e^{-t/\tau} \thetafct{t},
\end{equation}
where the Fourier transform reads
\begin{equation}
 \FT{h}\fargs{\omega}=\frac{1}{\left(1+\ci\tau\omega\right)^2}.
\end{equation}
This specific form is used for modeling synaptic interactions \cite{Bernard1994,Tuckwell_bookv1_1988}
and is usually called \emph{$\alpha$\nobreakdash-function} in a neuroscientific context.
We adopt this terminology in the following.
For the $\alpha$\nobreakdash-type response function, the elementary components $D_i$ of the integrand $I$ read
\begin{equation}\label{eq:Dalpha}
 \Da_i\fargs{\omega}=\frac{1}{\left(1+\ci\tau\omega\right)^2-\xi_i}.
\end{equation}
Similar to the previous section, some results are stated explicitly.

\paragraph{Two-point integral}
Given $\alpha$\nobreakdash-type response functions, the integral of Eq.~\eqref{eq:Iloop_int} reads
\begin{equation}\label{eq:Ii_j_alpha_int}
%  \mathcal{I}_{i,j}^\alpha\fargs{\omega_0} = \int_{-\infty}^\infty \frac{\intd \omega}{\left(\left(1+\ci\tau\left(\omega_0-\omega\right)\right)^2-\xi_{i}\right)\left(\left(1-\ci\tau\omega\right)^2-\xi_{j}\right)} .
\mathcal{I}_{i,j}^\alpha\fargs{\omega_0} = \int_{-\infty}^\infty 
 \Da_i\fargs{\omega_0-\omega}\Da_j\fargs{\omega} \intd \omega.
\end{equation}
The integrand $I_{i,j}^\alpha$ of $\mathcal{I}_{i,j}^\alpha$ has four poles,
\begin{align}
 a_\pm^{(i)} &=  \frac{\ci}{\tau}\left(\pm\sqrt{\xi_i}-1\right) +\omega_0, \\
 a_\pm^{(j)} &= -\frac{\ci}{\tau}\left(\pm\sqrt{\xi_j}-1\right).
\end{align}
The corresponding residues are given by
\begin{align}
 \residue{a_\pm^{(i)}}{I_{i,j}^\alpha} &= \pm \frac{\ci}{2\tau\sqrt{\xi_i}}\Da_j\fargs{a_\pm^{(i)}},\\
  \residue{a_\pm^{(j)}}{I_{i,j}^\alpha} &= \mp \frac{\ci}{2\tau\sqrt{\xi_j}}\Da_i\fargs{\omega_0-a_\pm^{(j)}}.
\end{align}
Summing up residues while assuming $|\myRe\fargs{\sqrt{\xi_i}}|<1$ for all $i\in\set{1,\dots,n}$ as discussed above, yields the analytical expression
\begin{equation}\label{eq:2pointIalpha}
 \mathcal{I}_{i,j}^\alpha\fargs{\omega_0} = \frac{\pi}{\tau\sqrt{\xi_j}} \left(\Da_i\fargs{\omega_0-a_{+}^{(j)}}-\Da_i\fargs{\omega_0-a_{-}^{(j)}}\right) .
\end{equation}

\paragraph{Three-point integral}

As for exponential response functions, a loop with three edges occurs in two variants
\begin{equation}\label{eq:Ii_jk_alpha_int}
%  \mathcal{I}_{i,jk}^\alpha\fargs{\omega_0,\nu_1} = \int_{-\infty}^\infty \frac{\intd \omega}{
%  \left(\left(1+\ci\tau\left(\omega_0-\omega\right)\right)^2-\xi_{i}\right)
%  \left(\left(1-\ci\tau\omega\right)^2-\xi_{j}\right)
%  \left(\left(1-\ci\tau\left(\nu_1+\omega\right)\right)^2-\xi_{j}\right)} .
 \mathcal{I}_{i,jk}^\alpha\fargs{\omega_0,\nu_1} = \int_{-\infty}^\infty 
 \Da_i\fargs{\omega_0-\omega} \Da_j\fargs{\omega} \Da_k\fargs{\nu_1+\omega} \intd \omega,
\end{equation}
and
\begin{equation}\label{eq:Iij_k_alpha_int}
 \mathcal{I}_{ij,k}^\alpha\fargs{\omega_0,\omega_1} = \int_{-\infty}^\infty 
 \Da_i\fargs{\omega_0-\omega} \Da_j\fargs{\omega_1-\omega} \Da_k\fargs{\omega} \intd \omega.
\end{equation}
The poles and residues are deduced analogously and are omitted at this point.

%%%%%%%%%%%%%%%%%%%%%%%%%%%%%%%%%%%%%%%%%%%%%%%%%%%%%%%%%%%
\subsection{\label{sec:explicit_preds}Explicit predictions}
%%%%%%%%%%%%%%%%%%%%%%%%%%%%%%%%%%%%%%%%%%%%%%%%%%%%%%%%%%%
The general loop integrals derived in Section~\ref{sec:genloopint} can now be used to obtain approximations for cumulants of any order. First we have a closer look at the motivating example from the beginning of this section. Afterwards, the one-loop correction to the second-order cumulant is calculated.

%%%%%%%%%%%%%%%%%%%%%%%%%%%%%%%%%%%%%%%%%%%%%%%%%%%%%%%%%%%%%%%%%%%%
\subsubsection{\label{sec:meancorr}One-loop correction to the rates}
%%%%%%%%%%%%%%%%%%%%%%%%%%%%%%%%%%%%%%%%%%%%%%%%%%%%%%%%%%%%%%%%%%%%

As stated earlier in Eq.~\eqref{eq:SMF_expansionpoint}, the working point $\bar r$ of our series expansion of the nonlinearity corresponds to the tree-level expectation value. This means
\begin{equation}
 \treeexpect{\delta z_i} = 0 \quad \forall i \in \set{1,\dots,n},
\end{equation}
and the working point is determined by solving the self-consistency equation
\begin{equation}\label{eq:workingpoint_selfconsistency}
 \bar r_i\fargs{t} = \phi\fargs{\sum_j\conv{h_{ij}}{\bar r_j}\fargs{t}+b_i\fargs{t}}.
\end{equation}

If we want to know how the tree-level covariances influence the rates, we have to calculate the one-loop correction
\begin{align}\nonumber
\ensexpect{\delta z_i}_\mathrm{1-loop} &=
\begin{picture}(70,40)(0,0)%\input{./../figures/tikz/FD_rate1loop.tikz}
    \begin{tikzpicture}[scale=1.,baseline=-.5ex,node distance=.8cm and 0.8cm]
    % draw vertices
    \node[outervertexstyle,label=above:$t$,label=below:{\scriptsize $i$}] (cl) {};
    \node[innervertexstyle,right=of cl,label=above:$t_1~$,label=below:{\scriptsize $j$}] (l4) {};
    \node[innervertexstyle,below right=of l4] (l1) {};
    \node[innervertexstyle,above right=of l4] (l3) {};
    \node[innervertexstyle,above right=of l1,label=above:$t_2$,label=below:{\scriptsize $l$}] (l2) {};
    % % Connect vertices with edges and draw weights
    \foreach \source/ \dest in {l4/cl,l2/l3,l2/l1}
        \draw[arrowedge] (\source) -- (\dest);
    \foreach \source/ \dest in {l3/l4,l1/l4}
        \draw[filteredge] (\source) -- (\dest);
    \end{tikzpicture}
\end{picture}\\\nonumber
% \vspace{20bp}
\\\nonumber
&= 2\cdot \sum_{j,l} \int \treeprop{ij}\fargs{t,t_1} \frac{\phi_j^{(2)}}{1!2!} \convfirst{h}{\treeprop{}}_{jl}\fargs{t_1,t_2} \convfirst{h}{\treeprop{}}_{jl}\fargs{t_1,t_2}\\\nonumber & \hspace*{72bp}\times\frac{\phi_l^{(0)}}{2!} \intd t_2 \intd t_1.
\end{align}
According to the Feynman rules in Section~\ref{sec:FeynmanRules}, there is only a single diagram contributing.
For a strict-sense stationary process, the propagator only depends on the time difference, $\treeprop{}\fargs{t_1,t_2}=\treeprop{}\fargs{t_1-t_2}$. 
This yields the Fourier representation
\begin{equation}\label{eq:prop_Fourierrep}
 \treeprop{}\fargs{t_1-t_2}=\frac{1}{2\pi}\int \FTtreeprop{}\fargs{\omega}e^{\ci\omega\left(t_1-t_2\right)}\intd\omega ,
\end{equation}
and thus
\begin{align}\nonumber\label{eq:convfirst_FT}
 \convfirst{h}{\treeprop{}}_{jl}\fargs{t_1-t_2} &=\frac{1}{2\pi}\int\sum_m\FT{h}_{jm}\fargs{\omega}\FTtreeprop{ml}\fargs{\omega}e^{\ci\omega\left(t_1-t_2\right)}\intd\omega\\
  &= \frac{1}{2\pi}\int E_{jl}\fargs{\omega}e^{\ci\omega\left(t_1-t_2\right)}\intd\omega.
 \end{align}
Using this Fourier representation of loop edges, one gets
\begin{align} \nonumber
\ensexpect{\delta z_i}_\mathrm{1-loop} &= \frac{1}{2}\frac{1}{\left(2\pi\right)^3}\int\sum_{j,l}
\FTtreeprop{ij}\fargs{\omega_1}E_{jl}\fargs{\omega_2}E_{jl}\fargs{\omega_3}\phi_j^{(2)}\phi_l^{(0)}\\
&\qquad\times e^{\ci\omega_1\left(t-t_1\right)} e^{\ci\omega_2\left(t_1-t_2\right)} e^{\ci\omega_3\left(t_1-t_2\right)}
\intd\omega_3\intd\omega_2\intd\omega_1\intd t_2\intd t_1 .
\end{align}
The time integrals yield $\delta$\nobreakdash-functions, such that
\begin{align}\nonumber
\ensexpect{\delta z_i}_\mathrm{1-loop} &= \frac{1}{4\pi}\int\sum_{j,l}
\FTtreeprop{ij}\fargs{\omega_1}E_{jl}\fargs{\omega_2}E_{jl}\fargs{\omega_3}\phi_j^{(2)}\phi_l^{(0)}\\\nonumber
&\qquad\times e^{\ci\omega_1 t}\deltafct{\omega_1-\omega_2-\omega_3} \deltafct{\omega_2+\omega_3}
\intd\omega_3\intd\omega_2\intd\omega_1\\
&= \frac{1}{4\pi}\int\sum_{j,l}
\FTtreeprop{ij}\fargs{0}E_{jl}\fargs{-\omega_3}E_{jl}\fargs{\omega_3} \phi_j^{(2)} \phi_l^{(0)}\intd\omega_3 .
\end{align}
In the last step, the $\omega_1$- and $\omega_2$\nobreakdash-integrations over the $\delta$\nobreakdash-functions are performed. 
With the shorthand from Eq.~\eqref{eq:Lloop_def}, the final result reads
\begin{equation}\label{eq:rate1loop}
 \ensexpect{\delta z_i}_\mathrm{1-loop} = \frac{1}{4\pi}\sum_{j,l} 
 \FTtreeprop{ij}\fargs{0} \mathcal{L}_{jl,jl}\fargs{0} \phi_j^{(2)} \phi_l^{(0)}.
\end{equation}
Thus, in total, the one-loop correction to the rates arising from the nonlinearity depends on the steady-state rate vector and the second derivative of the nonlinear gain function evaluated at the steady-state rate vector. Further, the simplest loop integral $\mathcal{L}_{jl,jl}\fargs{0}$ and the tree-level propagator contribute.

%%%%%%%%%%%%%%%%%%%%%%%%%%%%%%%%%%%%%%%%%%%%%%%%%%
\subsubsection{One-loop correction to the covariance}
%%%%%%%%%%%%%%%%%%%%%%%%%%%%%%%%%%%%%%%%%%%%%%%%%%
The covariance of the two spike trains $z_i$ and $z_j$ is given by
\begin{equation}
 C_{ij}\fargs{t+t^\prime,t^\prime} = \ensexpect{\delta z_i\fargs{t+t^\prime}\delta z_j\fargs{t^\prime}}.
\end{equation}
By means of the Feynman rules described in Section~\ref{sec:FeynmanRules}, the tree-level covariance can thus be deduced from the following diagram 
\begin{align}\nonumber
 C_{ij}^\mathrm{tree}\fargs{t+t^\prime,t^\prime} &= 
\begin{picture}(70,45)(0,0)%\input{./../figures/tikz/FD_treecov.tikz}
    \begin{tikzpicture}[scale=1.,baseline=-.5ex,node distance=.8cm and 0.8cm]
    % draw vertices
    \node (cl) {};
    \node[outervertexstyle,above=of cl,label=left:$t+t^\prime$,label=above:{\scriptsize $i$}] (l1) {};
    \node[innervertexstyle,below right=of l1,label=above:$t_1$,label=below:{\scriptsize $l$}] (c1) {};
    \node[outervertexstyle,below left=of c1,,label=left:$t^\prime$,label=below:{\scriptsize $j$}] (l2) {};
    % % Connect vertices with edges and draw weights
    \foreach \source/ \dest in {c1/l1,c1/l2}
        \draw[arrowedge] (\source) -- (\dest);
    \end{tikzpicture}
\end{picture} 
\\[30bp]
% &= \treeexpect{\delta z_i\fargs{t+t^\prime}\delta z_j\fargs{t^\prime}} \\
&= 2\cdot\sum_k\int \treeprop{ik}\fargs{t+t^\prime,t_1}\frac{\phi^{(0)}_k}{2!} \treeprop{jk}\fargs{t^\prime,t_1} \intd t_1 .
\end{align}
Considering again a strict-sense stationary process, the Fourier-transformed result reads
\begin{equation}\label{eq:treecov}
 \FT{C}_{ij}^\mathrm{tree}\fargs{\omega}=\sum_k \FTtreeprop{ik}\fargs{\omega}\phi^{(0)}_k\FTtreeprop{jk}\fargs{-\omega}.
\end{equation}
For linear mutually exciting point processes, this result matches the formula found by \cite{hawkesB_1971}.

In case of non-vanishing second or third derivative of the nonlinear gain function at the working point, the one-loop correction to the covariance is nonzero. 
First, all contributing terms have to be identified. The following two-step procedure yields all Feynman diagrams with one loop and two external vertices.
We start by creating all possible one-loop topologies, and then select the ones compatible with the Feynman rules from Section~\ref{sec:FeynmanRules}.
As the loop characteristic determines the topology, we start with a simple circle and successively add legs and internal propagators to it. An additional line can be attached to each internal propagator or internal vertex. Internal vertices are only added to internal propagators between two vertices with more than three incoming or outgoing legs. 
This procedure terminates after finitely many iterations, because the order of the cumulant of interest limits the number of legs added, and the number of internal propagators between two vertices with more than three incoming or outgoing legs is finite. Once all topologies are created, the ones relevant for the LNP cascade model are identified. This leads to a set of exactly 15 diagrams contributing to the second-order cumulant. The graphs whose contributions have to be calculated are given in Fig.~\ref{fig:FD_1loop_cov_2props}, Fig.~\ref{fig:FD_1loop_cov_3props}, and Fig.~\ref{fig:FD_1loop_cov_4props}. Individual contributions are labeled by $\mathcal{M}_{ij}^\beta$ with $\beta=1,\dots,15$.

The one-loop covariance correction is given by the sum over all contributions from the distinct diagrams
\begin{align}\label{eq:1loopcov}
 C_{ij}^{\mathrm{1-loop}}\fargs{t+t^\prime,t^\prime} &= \sum_{\beta=1}^{15} \mathcal{M}_{ij}^{\beta}\fargs{t+t^\prime,t^\prime}.
\end{align}
The authors of \cite{GKO_PILNP_2017} considered only a subset of these terms, namely $\mathcal{M}_{ij}^{\beta}$ for $\beta\in\set{1,\dots,5}$.
The remaining 10 contributions ($\beta=6,\dots,15$) were neglected by these authors.
However, these diagrams contain one loop and must be taken into account. Remember that the contributions of higher-order fluctuations are directly reflected by the number of loops in the diagrams \cite{Chow2015,GKO_PILNP_2017}.
The method used in \cite{GKO_PILNP_2017} to construct the loop diagrams for the covariance correction (cf.\ Fig.~13 in their article) was imperfect, though. 
An algorithm for the automated generation of Feynman diagrams is proposed in the discussion section of this article.

The individual contributions to the one-loop correction of the covariance in Eq.~\eqref{eq:1loopcov} are solved independently.
The procedure is illustrated for the $\mathcal{M}_{ij}^{1}$ component. For all other contributions we only list the result.

Translating the Feynman diagram into formula yields
\begin{align}\nonumber
\mathcal{M}_{ij}^{1}\fargs{t+t^\prime,t^\prime} &= \begin{picture}(90,40)(0,0)%\input{./../figures/tikz/FD_M11_detail.tikz}
    \begin{tikzpicture}[scale=1.,baseline=-.5ex,node distance=.8cm and 0.8cm]
    % draw vertices
    \node (cl) {};
    \node[outervertexstyle,above=of cl,label=left:$t+t^\prime$,label=above:{\scriptsize $i$}] (l1) {};
    \node[innervertexstyle,below right=of l1,label={[label distance=1ex]above:$t_1$},label=below:{\scriptsize $l$}] (c1) {};
    \node[outervertexstyle,below left=of c1,,label=left:$t^\prime$,label=below:{\scriptsize$j$}] (l2) {};
    \node[innervertexstyle,below right=of c1] (l) {};
    \node[innervertexstyle,above right=of c1] (u) {};
    \node[innervertexstyle,above right=of l,label=above right:$t_2$,label=below right:{\scriptsize $k$}] (c2) {};
    % % Connect vertices with edges and draw weights
    \foreach \source/ \dest in {c2/u,c2/l,c1/l1,c1/l2}
        \draw[arrowedge] (\source) -- (\dest);
    \foreach \source/ \dest in {l/c1,u/c1}
        \draw[filteredge] (\source) -- (\dest);
    \end{tikzpicture}
\end{picture}\vspace{20bp}
\\[1.5cm]\nonumber
  &= 4\cdot\int  \sum_{l,k} \treeprop{il}\fargs{t+t^\prime,t_1}\treeprop{jl}\fargs{t^\prime,t_1} \frac{\phi^{(2)}_l}{2!2!}\\
  &\qquad\qquad \times \convfirst{h}{\treeprop{}}_{lk}\fargs{t_1,t_2}  \convfirst{h}{\treeprop{}}_{lk}\fargs{t_1,t_2}\frac{\phi^{(0)}_k}{2!}
  \intd t_1\intd t_2 .
\end{align}

Considering again a strict-sense stationary process, the covariance only depends on the time lag $t$, and the reference time $t^\prime$ can be chosen arbitrarily.
Also the propagator depends only on the time difference and has the Fourier representation~\eqref{eq:prop_Fourierrep}.
% \begin{equation}
%  \treeprop{}\fargs{t_1-t_2}=\frac{1}{2\pi}\int \FTtreeprop{}\fargs{\omega}e^{\ci\omega\left(t_1-t_2\right)}\intd\omega.
% \end{equation}
Choosing $t^\prime=0$ and using Eq.~\eqref{eq:convfirst_FT}
% \begin{align}
%  \convfirst{h}{\treeprop{}}_{lk}\fargs{t_1-t_2} &=\frac{1}{2\pi}\int\sum_m\FT{h}_{lm}\fargs{\omega}\FTtreeprop{mk}\fargs{\omega}e^{\ci\omega\left(t_1-t_2\right)}\intd\omega\\
%   &= \frac{1}{2\pi}\int E_{lk}\fargs{\omega}e^{\ci\omega\left(t_1-t_2\right)}\intd\omega
%  \end{align}
results in 
\begin{align}\nonumber
 \mathcal{M}_{ij}^{1}\fargs{t} &= \frac{1}{2}\frac{1}{\left(2\pi\right)^4}\int\sum_{l,k}
 \FTtreeprop{il}\fargs{\omega_1}\FTtreeprop{jl}\fargs{\omega_2} E_{lk}\fargs{\omega_3}E_{lk}\fargs{\omega_4} \phi^{(2)}_l \phi^{(0)}_k \\
  &\qquad\times e^{\ci\omega_1\left(t-t_1\right)} e^{-\ci\omega_2 t_1} e^{\ci\omega_3\left(t_1-t_2\right)} e^{\ci\omega_4\left(t_1-t_2\right)}
 \intd\omega_4\intd\omega_3\intd\omega_2\intd\omega_1\intd t_2\intd t_1 .
\end{align}
The time integrals yield $\delta$\nobreakdash-functions such that
\begin{align}\nonumber
 \mathcal{M}_{ij}^{1}\fargs{t} &= \frac{1}{2}\frac{1}{\left(2\pi\right)^2}\int\sum_{l,k}
 \FTtreeprop{il}\fargs{\omega_1}\FTtreeprop{jl}\fargs{\omega_2} E_{lk}\fargs{\omega_3}E_{lk}\fargs{\omega_4} \phi^{(2)}_l \phi^{(0)}_k \\
  &\qquad\times e^{\ci\omega_1 t} \deltafct{\omega_1+\omega_2-\omega_3-\omega_4} \deltafct{\omega_3+\omega_4}
 \intd\omega_4\intd\omega_3\intd\omega_2\intd\omega_1 .
\end{align}
Executing the $\omega_2$- and $\omega_4$\nobreakdash-integrations leads to
\begin{align}\nonumber
 \mathcal{M}_{ij}^{1}\fargs{t} &= \frac{1}{2}\frac{1}{\left(2\pi\right)^2}\int\sum_{l,k}
 \FTtreeprop{il}\fargs{\omega_1}\FTtreeprop{jl}\fargs{-\omega_1} E_{lk}\fargs{\omega_3}E_{lk}\fargs{-\omega_3}\\\nonumber
 & \hspace*{72bp}\times \phi^{(2)}_l \phi^{(0)}_k e^{\ci\omega_1 t} \intd\omega_3\intd\omega_1 \\
 &=\frac{1}{2\pi}\int \frac{1}{4\pi}\sum_{l,k}\FTtreeprop{il}\fargs{\omega_1}\FTtreeprop{jl}\fargs{-\omega_1} \mathcal{L}_{lk,lk}\fargs{0} \phi^{(2)}_l \phi^{(0)}_k e^{\ci\omega_1 t}\intd\omega_1.
\end{align}
The contribution of the first diagram to the coherence between neuron $i$ and $j$ thus reads
\begin{equation}
\FT{\mathcal{M}}_{ij}^{1}\fargs{\omega} = \frac{1}{4\pi}\sum_{l,k}\FTtreeprop{il}\fargs{\omega}\FTtreeprop{jl}\fargs{-\omega} \mathcal{L}_{lk,lk}\fargs{0} \phi^{(2)}_l \phi^{(0)}_k
\end{equation}
where we use our loop integral abbreviations \eqref{eq:Lloop_def}.

The other contributions are derived similarly and the resulting terms read
\begin{flalign}
 \FT{\mathcal{M}}_{ij}^{2}\fargs{\omega} &= \frac{1}{4\pi}\sum_{l,k}
 \FTtreeprop{ik}\fargs{\omega}\FTtreeprop{jl}\fargs{-\omega}\mathcal{L}_{lk,lk}\fargs{-\omega} \phi^{(2)}_l \phi^{(0)}_k, &
\end{flalign}
\begin{flalign}
 \FT{\mathcal{M}}_{ij}^{3}\fargs{\omega} &= \frac{1}{4\pi}\sum_{l,k}
 \FTtreeprop{il}\fargs{\omega}\FTtreeprop{jk}\fargs{-\omega}\mathcal{L}_{lk,lk}\fargs{\omega} \phi^{(2)}_l \phi^{(0)}_k & \\\nonumber
 &= \FT{\mathcal{M}}_{ji}^{2}\fargs{-\omega} ,&
\end{flalign}
\begin{flalign}
 \FT{\mathcal{M}}_{ij}^{4}\fargs{\omega} &= \frac{1}{4\pi}\sum_{l,k,m}
 \FTtreeprop{il}\fargs{\omega}\FTtreeprop{jm}\fargs{-\omega}\mathcal{L}_{lk,lk}\fargs{\omega} E_{km}\fargs{\omega}\phi^{(2)}_l \phi^{(1)}_k \phi^{(0)}_m ,&
\end{flalign}
\begin{flalign}
 \FT{\mathcal{M}}_{ij}^{5}\fargs{\omega} &= \frac{1}{4\pi}\sum_{l,k,m}
 \FTtreeprop{im}\fargs{\omega}\FTtreeprop{jl}\fargs{-\omega}\mathcal{L}_{lk,lk}\fargs{-\omega} E_{km}\fargs{-\omega} \phi^{(2)}_l \phi^{(1)}_k \phi^{(0)}_m &\\\nonumber
 &= \FT{\mathcal{M}}_{ji}^{4}\fargs{-\omega} , &
\end{flalign}
\begin{flalign}
 \FT{\mathcal{M}}_{ij}^{6}\fargs{\omega} &= \frac{1}{4\pi}\sum_{l,k,m}
 \FTtreeprop{il}\fargs{\omega} \FTtreeprop{jk}\fargs{-\omega} \mathcal{L}_{lm,lkm}\fargs{0,\omega} \phi^{(2)}_l \phi^{(1)}_k \phi^{(0)}_m ,&
\end{flalign}
\begin{flalign}
 \FT{\mathcal{M}}_{ij}^{7}\fargs{\omega} &= \frac{1}{4\pi}\sum_{l,k,m}
  \FTtreeprop{ik}\fargs{\omega} \FTtreeprop{jl}\fargs{-\omega} \mathcal{L}_{lkm,lm}\fargs{-\omega,0} \phi^{(2)}_l \phi^{(1)}_k \phi^{(0)}_m ,&
\end{flalign}
\begin{flalign}
 \FT{\mathcal{M}}_{ij}^{8}\fargs{\omega} &= \frac{1}{4\pi}\sum_{m,l,k}
 \FTtreeprop{im}\fargs{\omega}\FTtreeprop{jm}\fargs{-\omega}E_{mk}\fargs{0}\mathcal{L}_{kl,kl}\fargs{0} \phi^{(1)}_m \phi^{(2)}_k \phi^{(0)}_l ,&
\end{flalign}
\begin{flalign}\label{eq:M9}
 \FT{\mathcal{M}}_{ij}^{9}\fargs{\omega} &= \frac{1}{4\pi}\sum_{k,l,m} 
 \FTtreeprop{im}\fargs{\omega}\FTtreeprop{jk}\fargs{-\omega} E_{km}\fargs{-\omega} \mathcal{L}_{kl,kl}\fargs{0} \phi^{(3)}_k \phi^{(0)}_l \phi^{(0)}_m, &
\end{flalign}
\begin{flalign}\label{eq:M10}
 \FT{\mathcal{M}}_{ij}^{10}\fargs{\omega} &= \frac{1}{4\pi}\sum_{k,l,m} 
 \FTtreeprop{ik}\fargs{\omega}\FTtreeprop{jm}\fargs{-\omega} E_{km}\fargs{\omega} \mathcal{L}_{kl,kl}\fargs{0} \phi^{(3)}_k \phi^{(0)}_l \phi^{(0)}_m &\\\nonumber
 &= \FT{\mathcal{M}}_{ji}^{9}\fargs{-\omega} ,&
\end{flalign}
\begin{flalign}\nonumber
 \FT{\mathcal{M}}_{ij}^{11}\fargs{\omega} &= \frac{1}{8\pi}\sum_{k,l,m,p}
  \FTtreeprop{ip}\fargs{\omega} \FTtreeprop{jk}\fargs{-\omega} \mathcal{L}_{klm,km}\fargs{-\omega,0} E_{lp}\fargs{-\omega} &\\
  &\hspace{72bp}\times\phi^{(2)}_k \phi^{(2)}_l \phi^{(0)}_m \phi^{(0)}_p,&
\end{flalign}
\begin{flalign}\nonumber
 \FT{\mathcal{M}}_{ij}^{12}\fargs{\omega} &= \frac{1}{8\pi}\sum_{k,l,m,p}
 \FTtreeprop{ik}\fargs{\omega} \FTtreeprop{jp}\fargs{-\omega} \mathcal{L}_{km,klm}\fargs{0,\omega} E_{lp}\fargs{\omega} &\\
  &\hspace{72bp}\times\phi^{(2)}_k \phi^{(2)}_l \phi^{(0)}_m \phi^{(0)}_p,&
\end{flalign}
\begin{flalign}\nonumber
 \FT{\mathcal{M}}_{ij}^{13}\fargs{\omega} &= \frac{1}{16\pi}\sum_{k,l,m,p}
  \FTtreeprop{ik}\fargs{\omega} \FTtreeprop{jp}\fargs{-\omega} E_{kp}\fargs{\omega}E_{kl}\fargs{0}
  \mathcal{L}_{lm,lm}\fargs{0} &\\
  &\hspace{72bp}\times\phi^{(2)}_k \phi^{(2)}_l \phi^{(0)}_m \phi^{(0)}_p,&
\end{flalign}
\begin{flalign}\nonumber
 \FT{\mathcal{M}}_{ij}^{14}\fargs{\omega} &= \frac{1}{16\pi}\sum_{k,l,m,p}
  \FTtreeprop{ip}\fargs{\omega} \FTtreeprop{jk}\fargs{-\omega} E_{kp}\fargs{-\omega}E_{kl}\fargs{0}
  \mathcal{L}_{lm,lm}\fargs{0} &\\
  &\hspace{72bp}\times\phi^{(2)}_k \phi^{(2)}_l \phi^{(0)}_m \phi^{(0)}_p &\\\nonumber
 &= \FT{\mathcal{M}}_{ji}^{13}\fargs{-\omega} ,&
\end{flalign}
\begin{flalign}\nonumber
 \FT{\mathcal{M}}_{ij}^{15}\fargs{\omega} &= \frac{1}{8\pi}\sum_{k,l,m,p}
  \FTtreeprop{ik}\fargs{\omega} \FTtreeprop{jl}\fargs{-\omega} \phi^{(2)}_k \phi^{(2)}_l \phi^{(0)}_m \phi^{(0)}_p &\\
   &\qquad\times \int E_{kp}\fargs{-\omega^\prime} E_{km}\fargs{\omega+\omega^\prime}E_{lm}\fargs{-\omega-\omega^\prime}E_{lp}\fargs{\omega^\prime} \intd \omega^\prime .&
\end{flalign}

For the comparison with data, it is convenient to work with the integrated cross-covariances
\begin{equation}
 c_{ij} = \int_{-\infty}^{\infty}C_{ij}\fargs{t} \intd t .
\end{equation}
Note that $c_{ij} = \FT{C}_{ij}\fargs{0}$ with $\FT{C}$ denoting the cross-spectra. We get this quantity directly by adding up all contributions from the previously listed terms evaluated at $\omega=0$.
The one-loop correction thus reads
\begin{align}\label{eq:intcov_loopcorrection}
 c_{ij}^{\mathrm{1-loop}} &= \sum_{\beta=1}^{15} \FT{\mathcal{M}}_{ij}^{\beta}\fargs{0} .
%  + \FT{\mathcal{M}}_{ij}^{00}\fargs{0} + \FT{\mathcal{M}}_{ij}^{10}\fargs{0} +
% 		\FT{\mathcal{M}}_{ij}^{01}\fargs{0} + \FT{\mathcal{M}}_{ij}^{20}\fargs{0} \\
% 		&\qquad + \FT{\mathcal{M}}_{ij}^{02}\fargs{0} + \FT{\mathcal{M}}_{ij}^{12}\fargs{0} +
% 		\FT{\mathcal{M}}_{ij}^{21}\fargs{0} + \FT{\mathcal{M}}_{ij}^{30}\fargs{0} + \FT{\mathcal{M}}_{ij}^{03}\fargs{0}.
\end{align}
For a prediction of the one-loop correction to the integrated covariance, the only loop integrals to be calculated from this general class are $\mathcal{L}_{lk,lk}\fargs{0}$,
$\mathcal{L}_{lm,lkm}\fargs{0,0}$, and $\mathcal{L}_{lkm,lm}\fargs{0,0}$.
% For vanishing frequencies, the two three-point integrals are equal. 
% 
% \section{\label{app:2pointL0}Specific two-point loop integral}
The simplest integral involved in the one-loop correction to the first- or second-order cumulant is $\mathcal{L}_{lk,lk}\fargs{0}$ which is solely based on $\mathcal{I}_{i,j}\fargs{0}$ from Eq.~\eqref{eq:Iloop_int}.
Let $\xi_{i}$ denote the $i$\nobreakdash-th eigenvalue of $D_{\phi^{(1)}}W$, as described previously.
Assuming for simplicity $\myRe{\xi_i}<1$ for all $i\in\set{1,\dots,n}$ (see remark in Section~\ref{sec:example_expdecay}),
the integral with exponentially decaying response function reads (cf.\ Eq.~\eqref{eq:Iij_exp})
\begin{equation}
 \mathcal{I}_{i,j}^\mathrm{e}\fargs{0}= \frac{2\pi}{\tau}\left(2-\xi_i-\xi_j\right)^{-1}.
\end{equation}
For an $\alpha$\nobreakdash-shaped response function, we assume $|\myRe\fargs{\sqrt{\xi_i}}|<1$ for all $i\in\set{1,\dots,n}$ and obtain from Eq.~\eqref{eq:2pointIalpha}
\begin{equation}
 \mathcal{I}_{i,j}^\alpha\fargs{0} = \frac{\pi}{\tau\sqrt{\xi_j}} \left(\Da_i\fargs{-a_{+}^{(j)}}-\Da_i\fargs{-a_{-}^{(j)}}\right),
\end{equation}
with $\Da_i\fargs{\omega}$ from Eq.~\eqref{eq:Dalpha}. This simplifies to
\begin{equation}
 \mathcal{I}_{i,j}^\alpha\fargs{0} = \frac{8\pi}{\tau} \left(\xi_j^2 + \left(\xi_i-4\right)^2-2\xi_j\left(4+\xi_i\right)\right)^{-1}.
\end{equation}
Putting everything together, we obtain
\begin{equation}
 \mathcal{L}_{lk,lk}^\mathrm{e}\fargs{0} = \frac{2\pi}{\tau} \sum_{i,j=1}^n \left(WV\right)_{li}V_{ik}^{-1} \left(WV\right)_{lj}V_{jk}^{-1} \left(2-\xi_i-\xi_j\right)^{-1}
\end{equation}
for the exponentially decaying response function, and 
\begin{equation}
 \mathcal{L}_{lk,lk}^\alpha\fargs{0} = \frac{8\pi}{\tau} \sum_{i,j=1}^n \left(WV\right)_{li}V_{ik}^{-1} \left(WV\right)_{lj}V_{jk}^{-1}
  \left(\xi_j^2 + \left(\xi_i-4\right)^2-2\xi_j\left(4+\xi_i\right)\right)^{-1}
\end{equation}
for the $\alpha$\nobreakdash-type response function.

% \section{\label{app:3pointL00}Specific three-point loop integrals}
For vanishing frequencies, the two three-point loop integrals $\mathcal{L}_{lm,lkm}\fargs{0,0}$ and $\mathcal{L}_{lkm,lm}\fargs{0,0}$ coincide. This can be seen by performing a transformation of the integration variable, $\omega \rightarrow -\omega^\prime$, in the integral $\mathcal{I}_{i_0 i_1, j_0}\fargs{0,0}$ from Eq.~\eqref{eq:Iloop_int}, which results in
\begin{equation}
\mathcal{I}_{i_0 i_1, j_0}\fargs{0,0} = \mathcal{I}_{j_0, i_0 i_1}\fargs{0,0}.
\end{equation}
If we assume again that all eigenvalues $\xi_i$ ($i=1,\dots,n$) of $D_{\phi^{(1)}}W$ have a real part strictly smaller than one, the integral with exponentially decaying response function reads
\begin{equation}
 \mathcal{I}^\mathrm{e}_{i_0 i_1, j_0} = \frac{2\pi}{\tau}\left(2-\xi_{j_0}-\xi_{i_0}\right)^{-1} \left(2-\xi_{j_0}-\xi_{i_1}\right)^{-1}.
\end{equation}
This results in 
\begin{align}\nonumber
 \mathcal{L}^\mathrm{e}_{lm,lkm}\fargs{0,0} = \mathcal{L}^\mathrm{e}_{lkm,lm}\fargs{0,0} &= \sum_{i_0,i_1,j_0=1}^n 
 \left(WV\right)_{li_0}V_{i_0k}^{-1}\left(WV\right)_{ki_1}V_{i_1m}^{-1}\\
 &\qquad\qquad\times\left(WV\right)_{lj_0}V_{j_0m}^{-1} \mathcal{I}^\mathrm{e}_{i_0 i_1, j_0}.
\end{align}
Assuming $|\myRe\fargs{\sqrt{\xi_i}}|<1$ and $\xi_i\neq 0$ for all $i\in\set{1,\dots,n}$, the integral with $\alpha$\nobreakdash-type response function is given by
\begin{align}\nonumber
 \mathcal{I}^\alpha_{i_0 i_1, j_0} = \frac{\pi}{\tau\sqrt{\xi_{j_0}}} & \left(\left(\left(2-\sqrt{\xi_{j_0}}\right)^2-\xi_{i_0}\right)^{-1}\left(\left(2-\sqrt{\xi_{j_0}}\right)^2-\xi_{i_1}\right)^{-1}\right.\\
 &\left.\quad - \left(\left(2+\sqrt{\xi_{j_0}}\right)^2-\xi_{i_0}\right)^{-1}\left(\left(2+\sqrt{\xi_{j_0}}\right)^2-\xi_{i_1}\right)^{-1}\right) .
\end{align}
This results in the complete loop integral
\begin{align}\nonumber
 \mathcal{L}^\alpha_{lm,lkm}\fargs{0,0} = \mathcal{L}^\alpha_{lkm,lm}\fargs{0,0} &= \sum_{i_0,i_1,j_0=1}^n 
 \left(WV\right)_{li_0}V_{i_0k}^{-1}\left(WV\right)_{ki_1}V_{i_1m}^{-1}\\
 &\qquad\qquad \times\left(WV\right)_{lj_0}V_{j_0m}^{-1} \mathcal{I}^\alpha_{i_0 i_1, j_0}. 
\end{align}

The one-loop correction $\mathcal{M}^{15}$ contains a loop integral differing from the general class of Eq.~\eqref{eq:Lloop_def}.
However, it can be reduced to an integral $\mathcal{I}_{i_0i_1,j_0j_1}$ from Eq.~\eqref{eq:Iloop_int}.
Note that one has
\begin{align}\label{eq:Kloop}\nonumber
 \int E_{kp}&\fargs{-\omega^\prime} E_{lm}\fargs{-\omega-\omega^\prime} E_{lp}\fargs{\omega^\prime}  E_{km}\fargs{\omega+\omega^\prime} \intd \omega^\prime = \\\nonumber
& \sum_{i_0,i_1,j_0,j_1}
 \left(WV\right)_{ki_0}V_{i_0p}^{-1} \left(WV\right)_{li_1}V_{i_1m}^{-1} \left(WV\right)_{lj_0}V_{j_0p}^{-1} \left(WV\right)_{kj_1}V_{j_1m}^{-1} \\
 &\qquad\times \underbrace{\int D_{i_0}\fargs{-\omega^\prime} D_{i_1}\fargs{-\omega-\omega^\prime} D_{j_0}\fargs{\omega^\prime} D_{j_1}\fargs{\omega+\omega^\prime} \intd \omega^\prime}_{\textstyle =\mathcal{I}_{i_0i_1,j_0j_1}\fargs{0,-\omega,\omega}} .
\end{align}
Thus a solution can be obtained in a similar manner by the residue calculus. 

To apply these results to the numerical simulations in Section~\ref{sec:numres_structLNP}, we need to calculate $\FT{\mathcal{M}}^{15}\fargs{0}$ for the $\alpha$\nobreakdash-shaped response function.
We thus need the analytic solution of $\mathcal{I}_{ij,kl}^\alpha\fargs{0,0,0}$.
The integral reads
\begin{equation}
 \mathcal{I}_{ij,kl}^\alpha\fargs{0,0,0} = \int \Da_i\fargs{-\omega} \Da_j\fargs{-\omega} \Da_k\fargs{\omega} \Da_l\fargs{\omega} \intd \omega,
\end{equation}
and assuming $|\myRe\fargs{\sqrt{\xi_i}}|<1$, the factors $\Da_k$ and $\Da_l$ cause poles of the integrand in the upper half-plane. Two cases have to be distinguished. We either have $\xi_k\neq\xi_l$ or $\xi_k=\xi_l$. In the former case, the relevant poles are
\begin{align}
 a^{(k)}_\pm &= -\frac{\ci}{\tau}\left(\pm\sqrt{\xi_k}-1\right),\\
 a^{(l)}_\pm &= -\frac{\ci}{\tau}\left(\pm\sqrt{\xi_l}-1\right),
\end{align}
and the residues of the integrand $I_{ij,kl}^\alpha$ are given by
\begin{align}
 \residue{a_\pm^{(k)}}{I_{ij,kl}^\alpha} &= \pm \frac{1}{2\ci\tau\sqrt{\xi_k}}\frac{1}{\xi_k-\xi_l}
      \left(\left(2\mp\sqrt{\xi_{k}}\right)^2-\xi_{i}\right)^{-1}\left(\left(2\mp\sqrt{\xi_{k}}\right)^2-\xi_{j}\right)^{-1},\\
 \residue{a_\pm^{(l)}}{I_{ij,kl}^\alpha} &= \pm \frac{1}{2\ci\tau\sqrt{\xi_l}}\frac{1}{\xi_l-\xi_k}
      \left(\left(2\mp\sqrt{\xi_{l}}\right)^2-\xi_{i}\right)^{-1}\left(\left(2\mp\sqrt{\xi_{l}}\right)^2-\xi_{j}\right)^{-1}.
\end{align}
In the latter case, where $\xi_l=\xi_k$, the poles $a^{(k)}_\pm = -\frac{\ci}{\tau}\left(\pm\sqrt{\xi_k}-1\right)$ are of order two. The residues are given by
\begin{align}\nonumber
 \residue{a_\pm^{(k)}}{I_{ij,kl}^\alpha} &= \frac{-\ci}{\tau} \left(\mp 4 +8\xi_k^{\frac{3}{2}}\mp\frac{5}{4}\xi_k^2
 +\sqrt{\xi_k}\left(16-2\xi_i-2\xi_j\right)
 \pm\xi_i\left(1-\frac{1}{4}\xi_j\right)\right.\\  \nonumber
 &\qquad\qquad\left.\mp\xi_k\left(18 - \frac{3}{4}\xi_i -\frac{3}{4}\xi_j \right) \pm\xi_j\right) \\
 & \qquad
 \times \xi_k^{-\frac{3}{2}}\left(\left(2\mp\sqrt{\xi_{k}}\right)^2-\xi_{i}\right)^{-2}\left(\left(2\mp\sqrt{\xi_{k}}\right)^2-\xi_{j}\right)^{-2}.
\end{align}
The final integral is given by summing up the residues according to Eq.~\eqref{eq:residuetheorem_I} and inserting the result into Eq.~\eqref{eq:Kloop} evaluated at $\omega=0$.

%%%%%%%%%%%%%%%%%%%%%%%%%%%%%%%%%%%%%%%%%%%%%%%%%%%%%%%%%%
\subsection{\label{sec:numres_structLNP}Numerical results}
%%%%%%%%%%%%%%%%%%%%%%%%%%%%%%%%%%%%%%%%%%%%%%%%%%%%%%%%%%
%%%%%%%%%%%%%%%%%%%%%%%%%
%% results here are from:
%% /home/kordovan/Simulations/data/structGLM/run_17-06-2019_05-00_MM_ER/analysis_0/
%%%%%%%%%%%%%%%%%%%%%%%%%%%%%%%%%%%%%%%%%%%%%%%%%%%%%%%%%%%%%%%%%%%%%%%%%%%%%%%%%%%%

This section provides a worked example for how loop corrections can improve cumulant predictions in the LNP cascade model. 
The results from Section~\ref{sec:explicit_preds} are compared to a simulation of the process. 
An outstanding feature of the theory are cumulant predictions for individual neurons, rather than population-averaged quantities.
We would like to stress this by looking at predictions for individual neurons or pairs of neurons
resulting in predictions of full distributions for the respective cumulants in the network, rather than just mean values.

%%%%%%%%%%%%%%%%%%%%%%%%%%%%%%%%%%%%
\subsubsection{Statistical measures}
%%%%%%%%%%%%%%%%%%%%%%%%%%%%%%%%%%%%

In order to compare our predictions, corresponding quantities have to be extracted from the simulated spike trains. 
In particular, the mean firing rate of a neuron is estimated by
\begin{equation}
 \check \lambda_i = \frac{\check{N}_i(\Delta)}{\Delta},
\end{equation}
where $\check{N}_i\fargs{\Delta}$ is the number of spikes in a time window of length $\Delta$.
Estimated quantities are marked with ``$~\check{~}~$'' throughout this section, to distinguish them from theoretical ones.
The integrated second-order cumulant $c_{ij}=\int_{-\infty}^{\infty}C_{ij}\fargs{t}\intd t=\FT{C}_{ij}\fargs{0}$, for simplicity denoted as covariance in the following, is estimated according to
\begin{equation}
 \check{c}_{ij}=\lim_{\Delta\to\infty} \frac{\covariance{\check{N}_i(\Delta)}{\check{N}_j(\Delta)}}{\Delta}.
\end{equation}
If we talk about cross-covariance in the following, we explicitly mean off-diagonal terms of this matrix ($i\neq j$). Auto-covariances are the diagonal elements ($i=j$).  

%%%%%%%%%%%%%%%%%%%%%%%%%%%%%%%%%%%%%
\subsubsection{Simulation parameters}
%%%%%%%%%%%%%%%%%%%%%%%%%%%%%%%%%%%%%

For numerical comparison, we simulate a two-population network of $N=250$ LNP neurons, which generate their spikes according to a Poisson distribution with conditional intensity (Eq.~\eqref{eq:intensity_LNP}),
\begin{equation}
\lambda_i\fargs{t} = \phi_i\fargs{\sum_j\conv{h_{ij}}{z_j}\fargs{t} +b_i\fargs{t}} .
\end{equation}
The excitatory population comprises $N_E=200$ and the inhibitory one $N_I=50$ neurons. 
The adjacency matrix is an \ER\ graph with connection probability $p=0.16$. The synaptic weights are given by $w_{EE}=0.12$ for excitatory-excitatory connections, $w_{IE} = 0.1$ for excitatory-to-inhibitory connections, and ${w_{EI}=w_{II}=-0.5}$ for all inhibitory connections.
The causal response function is chosen to be the $\alpha$\nobreakdash-function (defined in Eq.~\eqref{eq:alphafunc_def}) with synaptic time constant $\tau=10\millisec$.
Aiming for non-trivial loop corrections, the nonlinear gain function is a rectified quadratic function for all nodes
\begin{equation}
 \phi_i\fargs{x} = \thetafct{x}x^2 \qquad\forall i = 1,\dots,N.
\end{equation}
The baseline firing rate corresponding to $\phi_i\fargs{b}$ is set to $10\Hz$. The simulation time is $T=2\cdot 10^8\millisec$ with simulation time step of $\Delta t = 1\millisec$.
The simulation is implemented by the exact integration scheme outlined in \cite{Rotter1999}.
For data analysis, the first $10\,\mathrm{s}$ are dropped, and a spike count bin size of $\Delta_\mathrm{sc} = 1\,000\millisec$ is used. 
The simulation was implemented in Python using the open-source software SciPy \cite{SciPy}. 

%%%%%%%%%%%%%%%%%%%%%%%%%%%%%%%%
\subsubsection{Rate predictions}
%%%%%%%%%%%%%%%%%%%%%%%%%%%%%%%%

% \begin{figure}[tb]
%  \centering
% \subfloat[\label{fig:tree_ratecomp}Tree-level comparison.]{\includegraphics[width=.48\textwidth]{rates_comp_tree}}\quad%
% \subfloat[\label{fig:loop_ratecomp}One-loop-level comparison.]{\includegraphics[width=.48\textwidth]{rates_comp_1-loop}}
%  \caption{\label{fig:ratecomp}{\bf Rate comparisons for an \ER\ network.} Estimated rates versus tree-level \protect\subref{fig:tree_ratecomp} and one-loop \protect\subref{fig:loop_ratecomp} prediction.}
% \end{figure}

The simulated stationary rates are compared to the theoretical predictions in Fig.~\ref{fig:ratecomp}.
The tree-level prediction of the rates is given by $\bar r$ from Eq.~\eqref{eq:workingpoint_selfconsistency}, corresponding to $\treeexpect{\delta z}=0$. The one-loop correction is given by Eq.~\eqref{eq:rate1loop}, and the total one-loop prediction reads
\begin{equation}
 r_i^\mathrm{1-loop} = \bar r_i + \ensexpect{\delta z_i}_\mathrm{1-loop}.
\end{equation}
Figure~\ref{fig:ratecomp}(a) compares tree-level predictions, and a larger discrepancy is observed for increasing rates. The one-loop prediction perfectly corrects for this mismatch and we observe nearly perfect agreement in Fig.~\ref{fig:ratecomp}(b). 

The cumulative distribution function (cdf) of the rates is depicted in Fig.~\ref{fig:rates_cdf}. The mean tree-level (one-loop) prediction of the rates is $11.7\Hz$ ($12.5\Hz$), and the estimated mean rate is $12.6\Hz$. 
Excitatory neurons have a mean firing rate of $12.9\Hz$ (one-loop prediction), whereas for inhibitory ones the one-loop prediction is $10.8\Hz$. Both values are in good agreement with the estimated population-specific mean rates $13.0\Hz$ and $10.9\Hz$, respectively.
The relative one-loop corrections range from $3.8\,\%$ to $21.2\,\%$. 

% \begin{figure}%[htb]
%  \centering
%  \includegraphics[width=.8\textwidth]{rates_cdf}
% % \subfloat[\label{fig:corr_cdf}]{\includegraphics[width=.48\textwidth]{corr_cdf}}
%  \caption{\label{fig:rates_cdf}{\bf Cumulative distribution function of the rates in an \ER\ network.} Black dots and vertical lines indicate the mean of the distributions.
%  The dashed horizontal line at $0.5$ defines the median.}
% \end{figure}

To better quantify the discrepancy, the residuals, i.e.\ the difference between predicted and estimated rate, is investigated. Figure~\ref{fig:rate_residuals_cdf} depicts the cdf of the residuals for the tree- and loop-level prediction. 
For the two-population model under consideration, we further split into excitatory and inhibitory groups. The one-loop correction yields a relevant reduction of the error in all cases. 
While the residuals have a broad range (from $0.39\Hz$ to $1.87\Hz$) for tree-level predictions, the one-loop prediction corrects individual mismatches resulting in residuals ranging from $0.03\Hz$ to $0.13\Hz$ with a mean of $0.06\Hz$.  

% \begin{figure}%[htb]
%  \centering
%  \includegraphics[width=.8\textwidth]{rate_residuals_cdf}
%  \caption{\label{fig:rate_residuals_cdf}{\bf Residuals of rate predictions.} Cumulative distribution function for the residuals of the rate prediction in an \ER\ network. Purple lines are the residuals of the tree-level prediction and yellow ones are the residuals of the one-loop prediction. Residuals of only excitatory or only inhibitory neurons correspond to dashed and dotted lines, respectively. Black dots and vertical lines indicate the mean of the residuals. The dashed horizontal line at $0.5$ defines the median.}
% \end{figure}

%%%%%%%%%%%%%%%%%%%%%%%%%%%%%%%%%%%%%%
\subsubsection{Covariance predictions}
%%%%%%%%%%%%%%%%%%%%%%%%%%%%%%%%%%%%%%

Corrections arising from nonlinear gain functions are also calclulated for the second-order cumulant of the process. 
The tree-level prediction is given by ${c_{ij}^\mathrm{tree}=\FT{C}_{ij}^\mathrm{tree}\fargs{0}}$ from Eq.~\eqref{eq:treecov}.
For the full one-loop prediction, the one-loop correction $c_{ij}^{\mathrm{1-loop}}$ obtained by Eq.~\eqref{eq:intcov_loopcorrection} is added on top. 
Since the rectified quadratic nonlinearity $\phi$ has a vanishing third derivative, the contributions $\FT{\mathcal{M}}^{9}$ and $\FT{\mathcal{M}}^{10}$ are zero, cf.\ Eq.~\eqref{eq:M9} and Eq.~\eqref{eq:M10}.
All other terms yield nonzero corrections.

% \begin{figure}%[htb]
%  \centering
% \subfloat[\label{fig:crosscorr_cdf}]{\includegraphics[width=.8\textwidth]{crosscorr_cdf}}\par%
% \subfloat[\label{fig:autocorr_cdf}]{\includegraphics[width=.8\textwidth]{autocorr_cdf}}
%  \caption{\label{fig:autocross_cdfs}{\bf Cumulative distribution function of the covariances in an \ER\ network.} The cdfs of the cross-covariances (for distinct neurons $i\neq j$) are depicted in \protect\subref{fig:crosscorr_cdf} and the auto-covariances in \protect\subref{fig:autocorr_cdf}. Black dots and vertical lines indicate the mean of the distributions. The dashed horizontal line at $0.5$ defines the median.}
% \end{figure}

The mean of estimated cross-covariances is $0.4\Hz$, which is small as compared to the mean of estimated auto-covariances at $14.1\Hz$.
The predicted tree-level cdf, the one-loop cdf, and the empirical cdf of the cross-covariance, respectively, are shown in Fig.~\ref{fig:autocross_cdfs}(a), and the ones of the auto-covariance are depicted in Fig.~\ref{fig:autocross_cdfs}(b). 
A very basic observation is that the auto-covariances are much larger than the cross-covariances.
Because they roughly differ in one order of magnitude, we analyze the auto- and cross-covariance separately.
This observation provides valuable insight and can inform new beyond-mean-field models. 
In the simulated network setting, the cross-covariance has a multimodal distribution.
The one-loop correction improves the tree-level predictions for auto- and cross-covariances, as the one-loop cdf fits better to the estimated one.
In both cases, we observe an almost perfect match between the one-loop prediction and the distribution of empirical covariances. Remarkably, the entire distribution is precisely predicted, and not just its mean.

% \begin{figure}%[htb]
%  \centering
% \subfloat[\label{fig:crosscovres_cdf}]{\includegraphics[width=.48\textwidth]{crosscorr_residuals_cdf}}\quad%
% \subfloat[\label{fig:varres_cdf}]{\includegraphics[width=.48\textwidth]{var_residuals_cdf}}
%  \caption{\label{fig:cov_residuals_cdf}{\bf Residuals of covariance predictions.} The cdfs are shown separately for cross- and auto-covariances in \protect\subref{fig:crosscovres_cdf} and \protect\subref{fig:varres_cdf}, respectively. Black dots and vertical lines indicate the mean of the residuals. The dashed horizontal line at $0.5$ defines the median.}
% \end{figure}

For explicit comparison between simulation and prediction, the cdf of the residuals
for all possible pairs of distinct neurons are shown in Fig.~\ref{fig:cov_residuals_cdf}(a). 
While the distribution of cross-covariance residuals has a mean of $0.08\Hz$ with a standard deviation of $0.07\Hz$ for the tree-level prediction, the mean is $0.03\Hz$ for the one-loop predictions and the standard deviation slightly decreases to $0.04\Hz$.
In Fig.~\ref{fig:cov_residuals_cdf}(b) the residuals for auto-covariance predictions are shown. Here, the improvement achieved by the additional one-loop corrections is even more pronounced due to the larger absolute values of auto-covariances compared to cross-covariances. The tree-level residual distribution has a mean of $1.02\Hz$, whereas the mean of the one-loop residual distribution is $0.12\Hz$. As obvious from Fig.~\ref{fig:cov_residuals_cdf}(b), the standard deviation decreases drastically for residuals calculated with the one-loop predictions.
This decrease in standard deviation clearly illustrates that the one-loop covariance corrections are specific for each pair of neurons.
We stress once more that it is important to consider the entire distribution of residuals, because they can be either positive or negative. Therefore, the mean residual can be small, while the distribution spreads widely.

In total, we observe a remarkable improvement for second-order cumulant predictions. Both, the cross- and auto-covariance predictions get significantly closer to the respective quantities extracted from a simulation of the process, if one-loop diagram contributions are taken into account.

%%%%%%%%%%%%%%%%%%%%%%%%%%%%%%%%%%%%
\section{Conclusions and Discussion}
%%%%%%%%%%%%%%%%%%%%%%%%%%%%%%%%%%%%
%%%%%%%%%%%%%%%%%%%%%%%%%%%%%%%%%%%%

Predicting spike train cumulants in nonlinear spiking models is complicated by the fact that higher-order moments are coupling to lower-order ones. 
One approach is to expand the moment hierarchy in a series, where higher-order terms account for the influence of higher-order moments on the dynamics.
Specifically, when representing the spiking activity in LNP cascade models by path integrals and appropriate path densities,
one obtains a systematic series expansion for spike train cumulants.
This series is called \emph{loop expansion}, as the number of loops in the Feynman diagrams associated with the terms in the expansion reflect the successive contributions of higher-order fluctuations. 
For actual predictions the series expansion is truncated, ignoring contributions beyond the cutoff.
In this work, the truncation happens at the one-loop level. While tree-level predictions are independent of any influences by higher-order cumulants, the one-loop correction to a cumulant of order $n$ accounts for the tree-level contribution from the $(n+1)$\nobreakdash-order cumulant. 
We justify this truncation for specific network models by comparing it to numerical simulations.  
A more general justification can be derived, if the loop contributions can be related to a small parameter like the inverse network size \cite{bressloff_systemsize_2009}. 

For parameter regimes where the truncation is meaningful, 
corrections to spike train cumulants can be calculated using the loop expansion.
We found a way to make these predictions more robust by analytically solving the loop integrals.

As calculations are very extensive even for low-order cumulants, a systematic implementation using computer algebra is indispensable.
This can be done, for example, as in theoretical particle physics using programs like \textsc{FeynArts} \cite{hahn_feynarts_2001} and \textsc{FormCalc} \cite{hahn_formcalc_1999},
which are written in \texttt{mathematica}\textsuperscript{\textregistered}. 
All contributing graph topologies have to be generated first. 
The procedure used for the graph generation can be directly implemented, but a mathematically rigorous proof of completeness is still missing.
This is essential in order not to forget any term of a given order in the series expansion, as it has happened previously for seemingly simple corrections of second-order cumulants \cite{GKO_PILNP_2017}.
Individual graphs are then translated into formulas by making use of the Feynman rules for the stochastic process under consideration.
A cascade of calculations is then performed on these expressions by making use of the new results explained above.
Loop integrals can be replaced by sums over residuals, which can be calculated in a standardized way.
Having such a machinery at hand allows to easily obtain more accurate predictions for stationary spiking statistics.

Based on the theoretical description of a stochastic process, the approach presented here
makes quantitative and testable predictions for spike train statistics.
Specifically, we were able to compute second-order cumulants for all individual tuples of neurons. 
As a result, full distributions of spike train cumulants, and not just their mean and variance, can be computed for given neuronal populations. 
Although the corrections due to an instantaneous nonlinearity are small, they can be nevertheless important. 
Spike-timing-dependent structural plasticity, for example, depends on the second-order cumulant of activity \cite{Deger_STDStructP_2012}.
Small accumulating errors in theoretical predictions can cause large discrepancies of the predicted network structure.
In homeostatic structural plasticity \cite{Gallinaro2018}, in contrast,
the degree of each node depends on its activity working point in a nonlinear manner.
Small discrepancies in the working point prediction yields wrong degree predictions. 
Especially for non-normally distributed firing rates, like for example heavy-tailed log-normal distributions, it is important to know the entire distribution and not just its first two moments. 

A final remark concerns the magnitude of individual diagram contributions. Although it is consistent to take all diagrams with a given number of loops into account, these do not automatically provide corrections of comparable magnitude. Depending on the parameters, specific diagrams can make much stronger contributions than others. The size of the contribution depends, for example, on the number of internal filter edges, or different occurrences of nonlinear gain function derivatives. Additionally, single diagrams may specifically correct cumulant predictions for certain subsets of neurons. A systematic investigation of these individual contributions would provide further insight into structure-dynamics relations.
For future analyses of this type, it comes handy to have the explicit analytical dependencies of the loop integrals available. 
In total, these investigations greatly advance our understanding of how nonlinearities shape the cumulant distributions in networks. 

\section*{Abbreviations}

cdf, cumulative distribution function; LNP, linear-nonlinear Poisson; pdf, probability density functional; STDP, spike-timing dependent plasticity

\section*{Declarations}
%%%%%%%%%%%%%%%%%%%%%%%%%%%%%%%%%%%%%%%%%%%%%%
%%                                          %%
%% Backmatter begins here                   %%
%%                                          %%
%%%%%%%%%%%%%%%%%%%%%%%%%%%%%%%%%%%%%%%%%%%%%%

\begin{backmatter}

\section*{Ethics approval and consent to participate}
  Not applicable.
\section*{Consent for publication}
  Not applicable.
\section*{Availability of data and material}
  The simulation software scripts and generated data used and analysed during the current study are available from the corresponding author on reasonable request.
\section*{Competing interests}
  The authors declare that they have no competing interests.
\section*{Funding}
  The authors acknowledge support by the state of Baden-Württemberg through bwHPC and the German Research Foundation (DFG) through grant no INST 39/963-1 FUGG (bwForCluster NEMO). The research reported here has received additional funding from the European Union’s Seventh Framework Programme (FP7/2007-2013) under Grant Agreement 600925 (NeuroSeeker), from BrainLinks-BrainTools, Cluster of Excellence funded by the German Research Foundation (DFG, grant number EXC 1086), and from the Carl Zeiss Stiftung.
\section*{Authors' contributions}
  The study was conceived and designed by MK and SR. Simulations, calculations and analysis of data was performed by MK. MK and SR drafted and revised the manuscript of the publication and approved its final version.
\section*{Acknowledgements}
  The authors are grateful to \emph{Stojan Jovanovi\'{c}}, \emph{Christopher Kim}, and \emph{Neboj\v{s}a Ga\v{s}parovi\'{c}} for their advice and helpful discussions.
  
%%  After submission of the .TEX file,                     %%
%%  you will be prompted to submit your .BBL file.         %%

% if your bibliography is in bibtex format, use those commands:
\bibliographystyle{bmc-mathphys} % Style BST file (bmc-mathphys, vancouver, spbasic).
\bibliography{CumulantsLNP_PathIntegral}      % Bibliography file (usually '*.bib' )
% for author-year bibliography (bmc-mathphys or spbasic)
% a) write to bib file (bmc-mathphys only)
% @settings{label, options="nameyear"}

% or include bibliography directly:
% \begin{thebibliography}
% \bibitem{b1}
% \end{thebibliography}

%%%%%%%%%%%%%%%%%%%%%%%%%%%%%%%%%%%
%%                               %%
%% Figures                       %%
%%                               %%
%% NB: this is for captions and  %%
%% Titles. All graphics must be  %%
%% submitted separately and NOT  %%
%% included in the Tex document  %%
%%                               %%
%%%%%%%%%%%%%%%%%%%%%%%%%%%%%%%%%%%

%%
%% Do not use \listoffigures as most will included as separate files

\section*{Figures}
  \begin{figure}[h!]
    \includegraphics{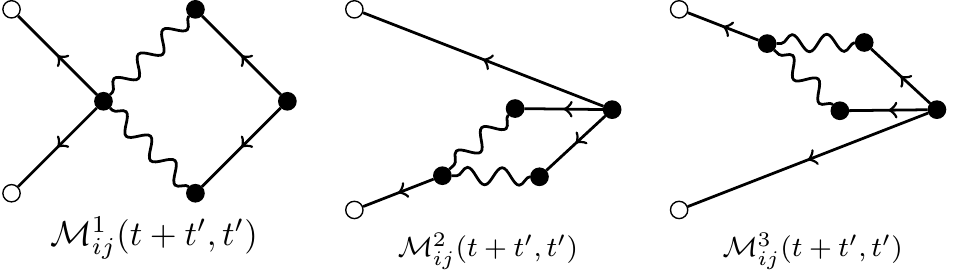}
    \caption{\label{fig:FD_1loop_cov_2props}\csentence{Feynman diagrams 1.}
    Diagrams with two internal filter edges contributing to the one-loop correction for the covariances.} 
  \end{figure}

  \begin{figure}[h!]
    \includegraphics{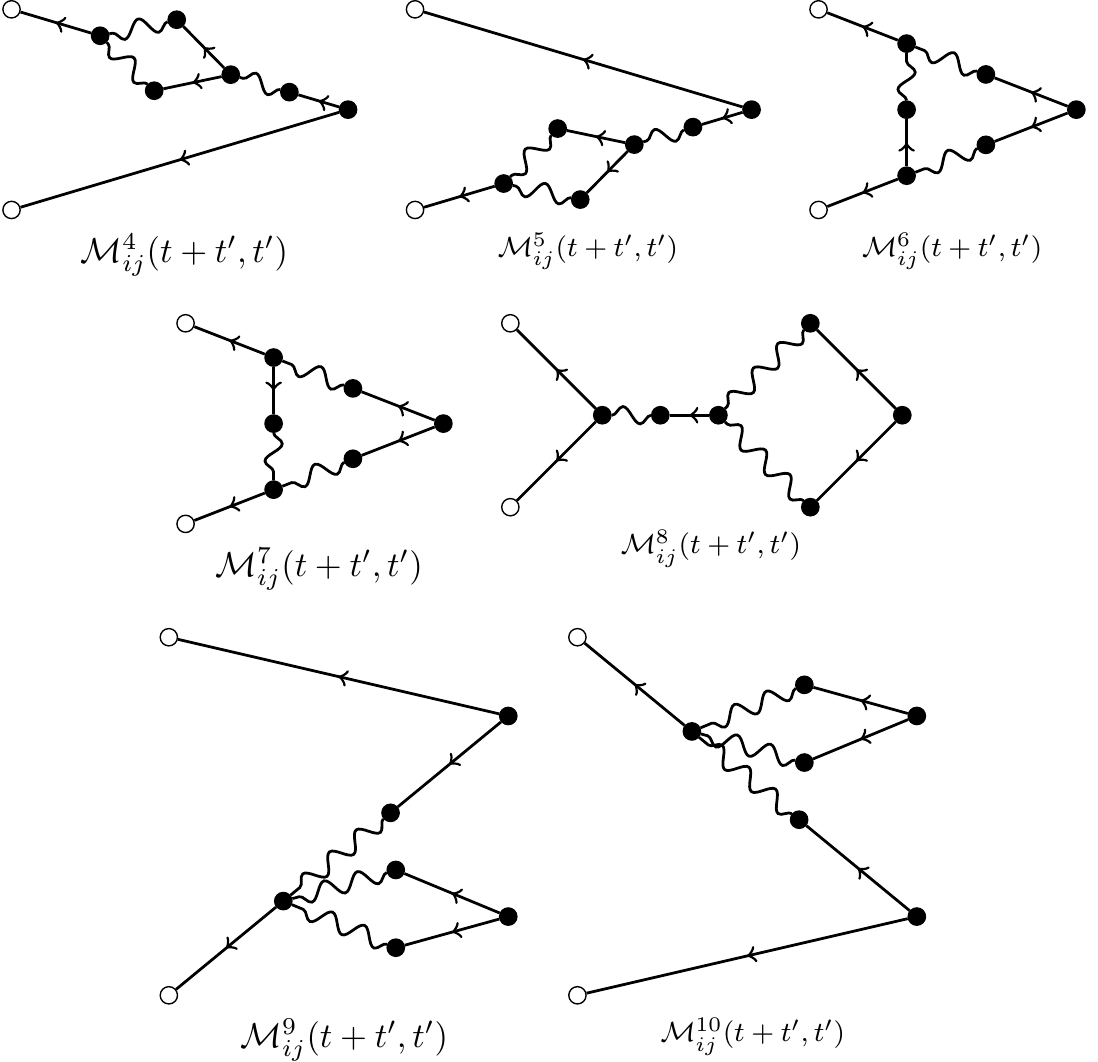}
    \caption{\label{fig:FD_1loop_cov_3props}\csentence{Feynman diagrams 2.} 
    Diagrams with three internal filter edges contributing to the one-loop correction for the covariances.}   
  \end{figure}
  
  \begin{figure}[h!]
   \includegraphics{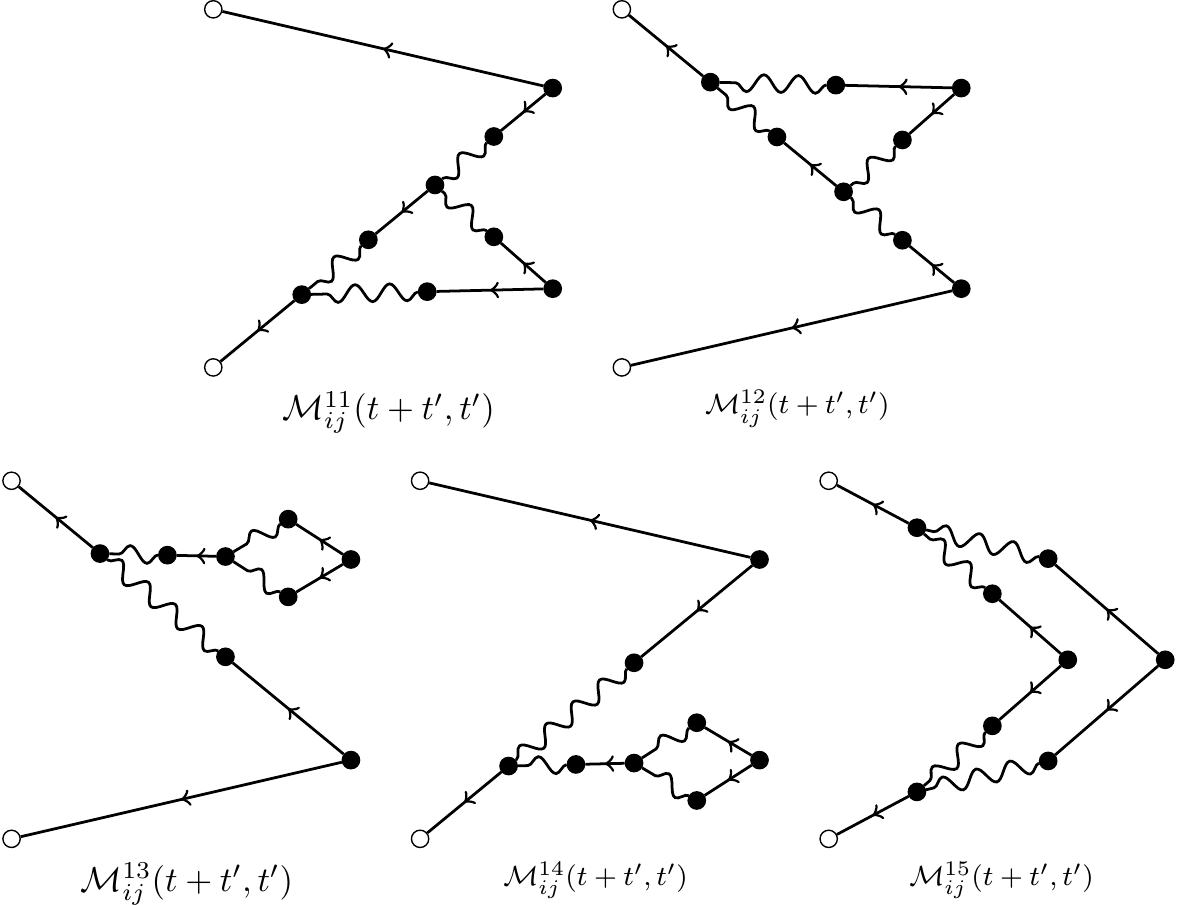}
   \caption{\label{fig:FD_1loop_cov_4props}\csentence{Feynman diagrams 3.}
   Diagrams with four internal filter edges contributing to the one-loop correction for the covariances.} 
  \end{figure}

  \begin{figure}[h!]
   \includegraphics{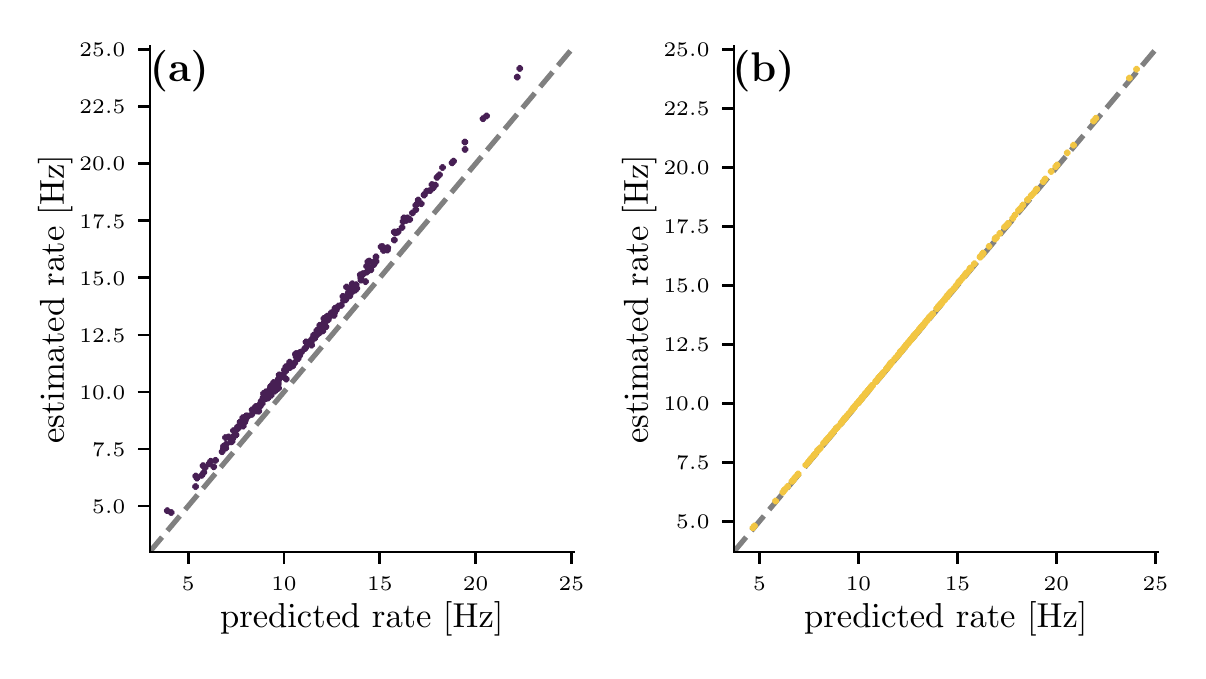}
  \caption{\label{fig:ratecomp}\csentence{Rate comparisons for an \ER\ network.} Estimated rates versus tree-level (a) and one-loop (b) prediction.}
  \end{figure}

  \begin{figure}[h!]
   \includegraphics{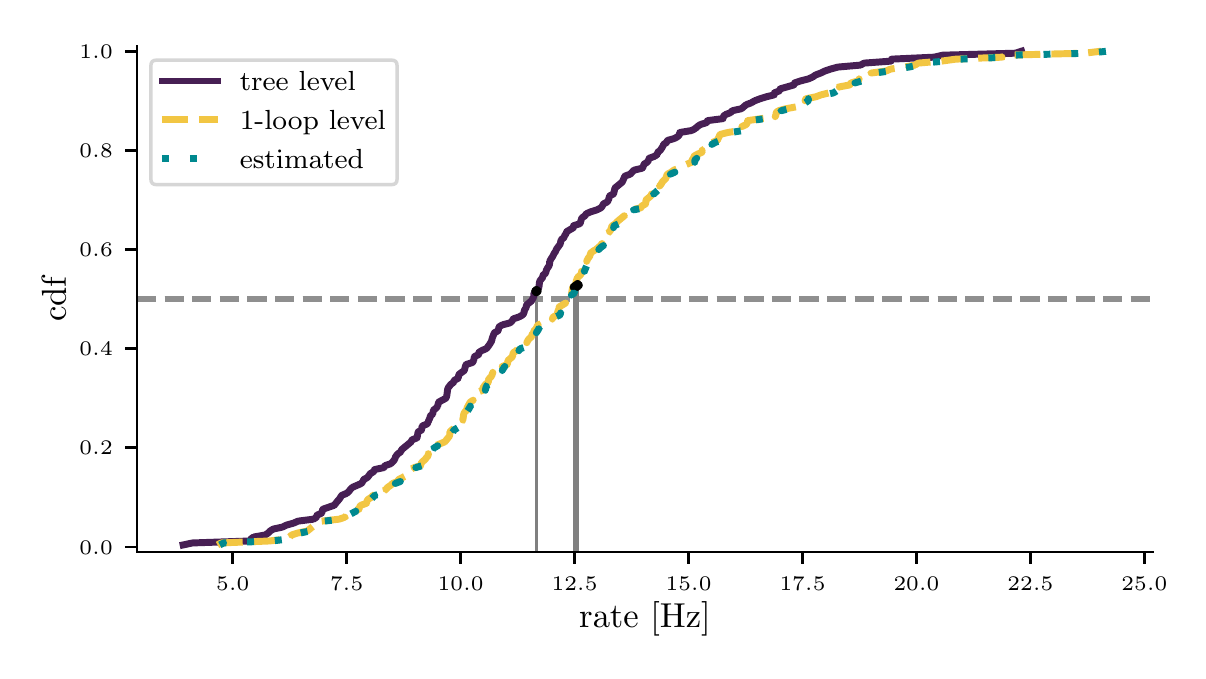}
    \caption{\label{fig:rates_cdf}\csentence{Cumulative distribution function of the rates in an \ER\ network.} Black dots and vertical lines indicate the mean of the distributions.
 The dashed horizontal line at $0.5$ defines the median.}
  \end{figure}

  \begin{figure}[h!]
   \includegraphics{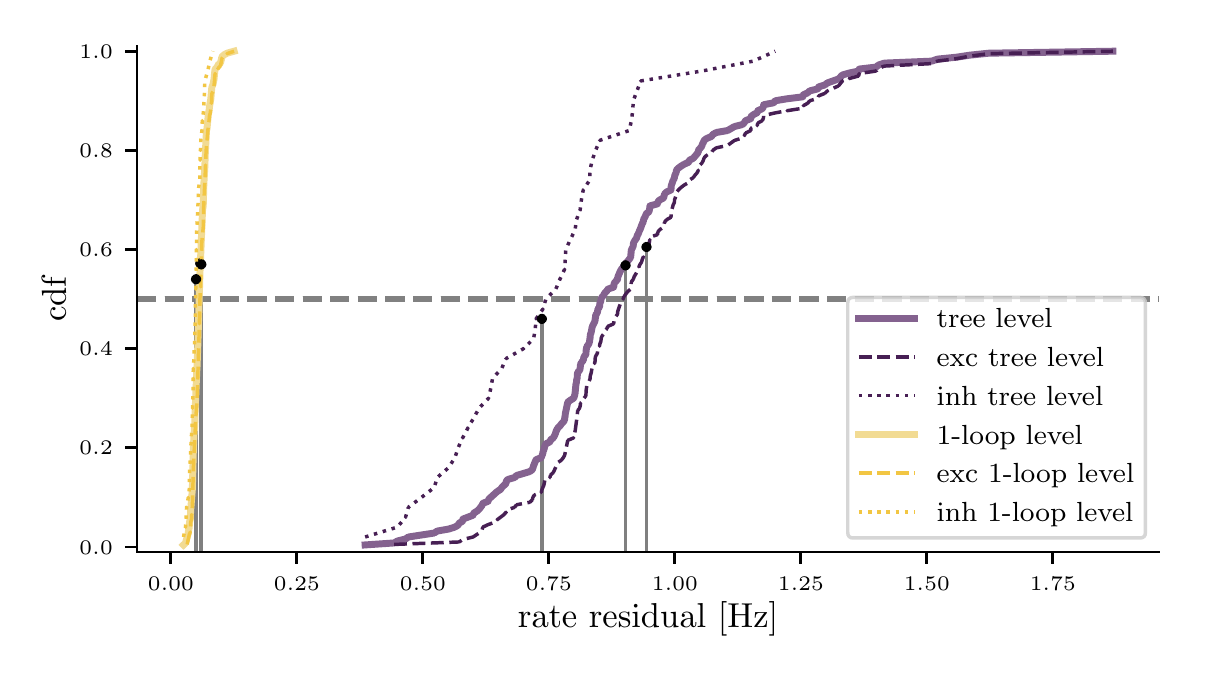}
    \caption{\label{fig:rate_residuals_cdf}\csentence{Residuals of rate predictions.} Cumulative distribution function for the residuals of the rate prediction in an \ER\ network. Purple lines are the residuals of the tree-level prediction and yellow ones are the residuals of the one-loop prediction. Residuals of only excitatory or only inhibitory neurons correspond to dashed and dotted lines, respectively. Black dots and vertical lines indicate the mean of the residuals. The dashed horizontal line at $0.5$ defines the median.}
  \end{figure}

  \begin{figure}[h!]
   \includegraphics{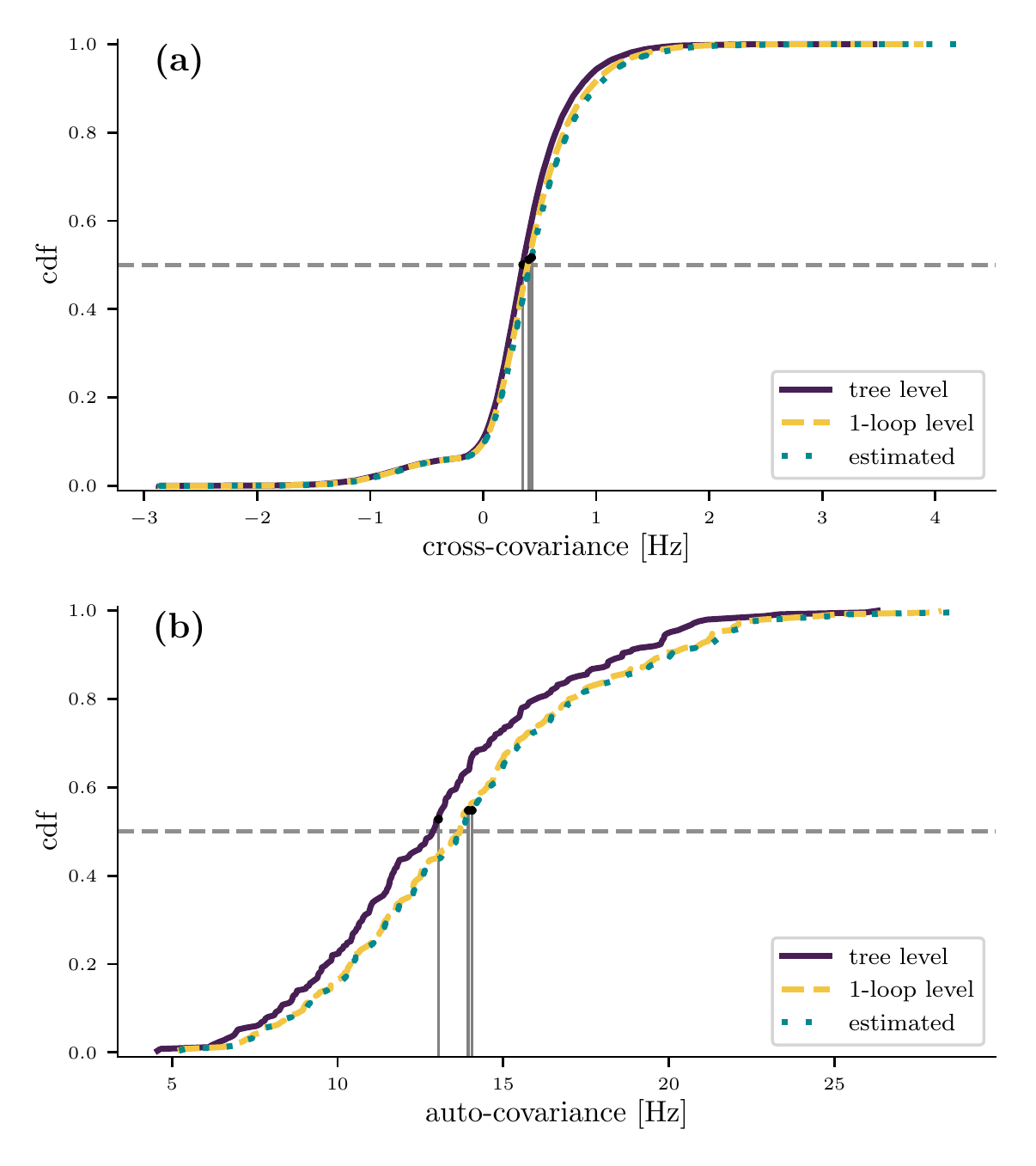}
    \caption{\label{fig:autocross_cdfs}\csentence{ Cumulative distribution function of the covariances in an \ER\ network.} The cdfs of the cross-covariances (for distinct neurons $i\neq j$) are depicted in (a) and the auto-covariances in (b). Black dots and vertical lines indicate the mean of the distributions. The dashed horizontal line at $0.5$ defines the median.}
  \end{figure}

  \begin{figure}[h!]
   \includegraphics{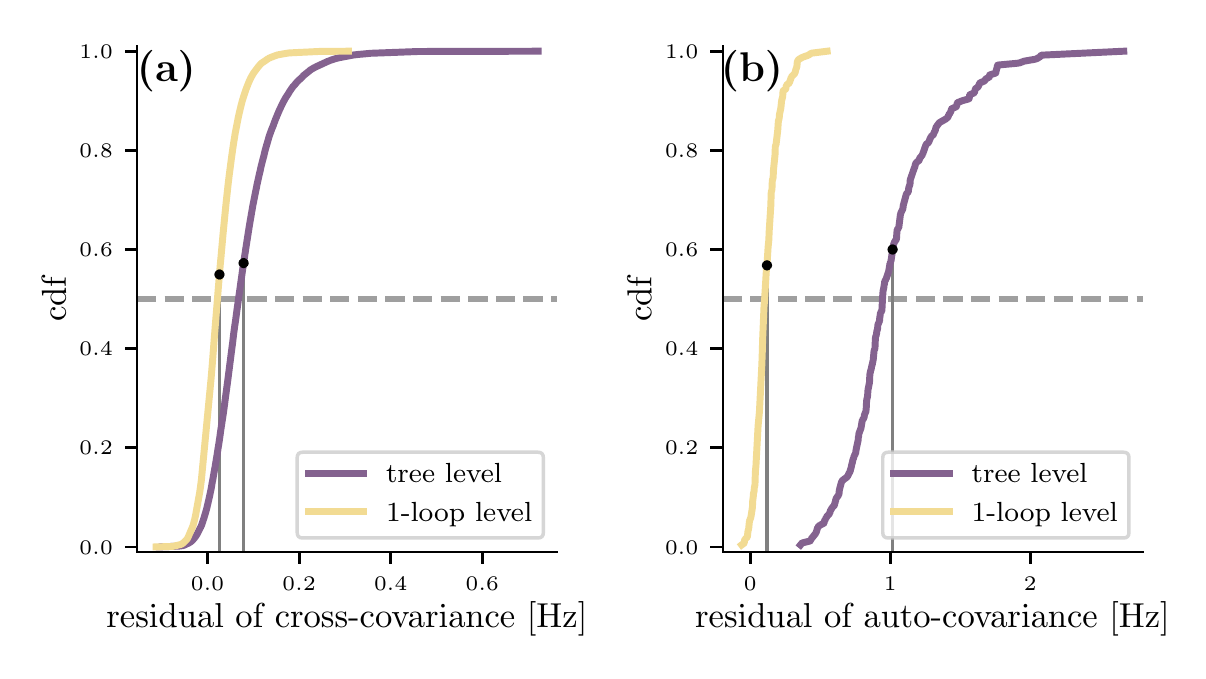}
    \caption{\label{fig:cov_residuals_cdf}\csentence{Residuals of covariance predictions.} The cdfs are shown separately for cross- and auto-covariances in (a) and (b), respectively. Black dots and vertical lines indicate the mean of the residuals. The dashed horizontal line at $0.5$ defines the median.}
  \end{figure}

%%%%%%%%%%%%%%%%%%%%%%%%%%%%%%%%%%%
%%                               %%
%% Tables                        %%
%%                               %%
%%%%%%%%%%%%%%%%%%%%%%%%%%%%%%%%%%%

%% Use of \listoftables is discouraged.
%%
% \section*{Tables}
% \begin{table}[h!]
% \caption{Sample table title. This is where the description of the table should go.}
%       \begin{tabular}{cccc}
%         \hline
%            & B1  &B2   & B3\\ \hline
%         A1 & 0.1 & 0.2 & 0.3\\
%         A2 & ... & ..  & .\\
%         A3 & ..  & .   & .\\ \hline
%       \end{tabular}
% \end{table}

%%%%%%%%%%%%%%%%%%%%%%%%%%%%%%%%%%%
%%                               %%
%% Additional Files              %%
%%                               %%
%%%%%%%%%%%%%%%%%%%%%%%%%%%%%%%%%%%

% \section*{Additional Files}
%   \subsection*{Additional file 1 --- Sample additional file title}
%     Additional file descriptions text (including details of how to
%     view the file, if it is in a non-standard format or the file extension).  This might
%     refer to a multi-page table or a figure.
% 
%   \subsection*{Additional file 2 --- Sample additional file title}
%     Additional file descriptions text.

%\listoftodos

\end{backmatter}
\end{document}